\documentclass[a4paper,12pt]{article}
\usepackage[english]{babel}
\usepackage{amsmath, amssymb,graphicx}
\usepackage{jheppub}

\newcommand{\be}{\begin{equation}}
\newcommand{\ee}{\end{equation}}
\newcommand{\bea}{\begin{eqnarray}}
\newcommand{\eea}{\end{eqnarray}}
\newcommand{\nn}{\nonumber}

\newcommand{\uu}{\bf{u}}
\newcommand{\ww}{\bf{w}}

\title{Holographic Dual to Conical Defects: II. Colliding Ultrarelativistic Particles}
\author[a]{D.S. Ageev}
\author[a]{I.Ya. Aref'eva}
\affiliation[a]{Steklov Mathematical Institute, Russian Academy of Sciences, Gubkin str. 8, 119991
Moscow, Russia}

\emailAdd{ageev@mi.ras.ru}
\emailAdd{arefeva@mi.ras.ru}

\abstract{We study instant conformal symmetry breaking as a holographic effect of ultrarelativistic particles moving in the $AdS_3$ spacetime. We give the qualitative picture of this effect probing it by two-point correlation functions and the entanglement entropy of the corresponding boundary theory.
We show that within geodesic approximation
the ultra-relativistic massless  defect
due to gravitational lensing of the geodesics,  produces a zone structure for correlators with broken conformal invariance. Meanwhile, the holographic entanglement entropy also exhibits a transition to the non-conformal behaviour. Two colliding massless defects produce more
diverse zone structure for correlators and  the entanglement entropy.}


\begin{document}
\maketitle

\section{Introduction}

In this paper we continue to study
two-dimensional quantum field theory on the boundary of the $AdS_3$ space deformed by point particles moving
in the bulk within the $AdS/CFT$ correspondence.  In the previous paper
\cite{AAT} we have studied deformations by massive moving particles
in $AdS_3$.  Similar problems have been investigated in the early papers
   \cite{Balasubramanian:1999zv,AB-TMF, Balasubramanian:2014sra,Arefeva:2015sza} for various models.  For motivation to study these problems see  \cite{AAT} and references therein.

    In  papers \cite{Balasubramanian:1999zv,AB-TMF, Balasubramanian:2014sra,Arefeva:2015sza},
    \cite{AAT} the group theoretical language is used to describe the conical defects \cite{Deser,Hooft,Matschull:1998rv} by the corresponding cutting and gluing procedure. To calculate the two-point boundary correlators we use the geodesic approximation proposed in this context in \cite{Balasubramanian:1999zv}.
The ultrarelativistic point particle, starting from the boundary of the cylinder shrinks the bulk along the worldline symmetrically with respect to the starting point \cite{Matschull:1998rv}. As the particle penetrates the bulk of  $AdS_3$ deeper geodesics connecting the boundary points exhibit  lensing effects. A similar effect takes place for massive particles  \cite{Balasubramanian:1999zv,AB-TMF, Balasubramanian:2014sra,Arefeva:2015sza},
    \cite{AAT} and due to this effect one gets the zone structure for two-point correlators of the boundary theory. Considering collisions of two ultrarelativistic particles we get even more complicated structure for geodesics, that lead to a multi-zone structure for two-point correlators. Namely, around the edges of the shrinking space we get the focusing of geodesics due to winding on the wedges of the defect in a rather  nontrivial way. Near the endpoints the winding geodesics dominate, while away from the location of wedges, dominate the non-winding ones.
    There is also an intermediate zone, where both families of geodesics contribute, creating some kind of resonance. There are also discontinuities separating  different zones, they are  localized
    and  propagate on the boundary of  $AdS_3$ with the constant speed.

 The study of the holographic entanglement entropy (HEE) \cite{Ryu:2006bv} becomes a rapidly developing subject with
a broad range of applications due relative simplicity of  entanglement entropy realization in holography and wide variety
of modifications of basic examples  \cite{Nishioka:2009un,Asplund:2014coa,Nozaki:2013wia,Marolf:2015vma,
 Casini:2015zua,Hartman:2015apr}. In this paper we calculate the HEE for 2-dimensional theories on the
circle with varying radius. As  holographical gravity models we consider the same models as in the first part of the paper,   the $AdS_3$ spacetime deformed by one or  two ultrarelativistic particles.

The paper is organized as follows. In Section 2 we review the $AdS_3/CFT_{2}$ setup   with colliding point particles in
the bulk.  In Section 3 we compute the two-point correlation function using the geodesic approximation and present the results of the calculations.  In Section 4 we compute the holographic entanglement entropy in  presence of ultra relativistic particles. In  the conclusion we summarize the obtained results and discuss future perspectives related with investigations of collisions of two ultrarelativistic particles.

\section{Setup}
Let us briefly recall the group structure of the description of ultrarelativistic particle deformation of the $AdS_3$ space on group theoretical language \cite{Deser,Hooft,Matschull:1998rv}.

Points of $AdS_3$  can be represented as $SL(2)$ group elements of real $2\times 2$ matrices
\be
{\bf x} = x_3{\bf 1}+\sum _{\mu=0,1,2} \gamma_\mu x^\mu
=\cosh \chi\,{\bf \Omega}(t)+\sinh\chi\,{\bf \Gamma}(\phi),
\label{x-br}\ee
where
\begin{equation}
{\bf 1} =\left(
           \begin{array}{cc}
             1 & 0 \\
             0 & 1 \\
           \end{array}
         \right)
;~~~~
\gamma_0 = \left(
             \begin{array}{cc}
               0 & 1 \\
               -1 & 0 \\
             \end{array}
           \right)
;~~~~
\gamma_1 = \left(
             \begin{array}{cc}
               0 & 1 \\
               1 & 0 \\
             \end{array}
           \right)
;~~~~
\gamma_2 =\left(
            \begin{array}{cc}
              1 & 0 \\
              0 & -1 \\
            \end{array}
          \right),
\end{equation}
and
\be
{\bf \Omega}(t)=\cos t {\bf 1}+\sin t\, \gamma_0\,\,\,\,\,\,
{\bf \Gamma}(\phi)=\cos\phi \gamma_1+\sin\phi \gamma_2,\ee
where $(t,\chi,\phi)$ are "barrel" coordinates, here we assume, that
$\chi\geq 0$, $\phi\simeq \phi+2\pi$, $-\pi/2<t<\pi/2$.

We will also use the Poincare disk coordinate
$r$, related with $\chi$ via
\bea
\label{r-tanh}
r&=&\tanh(\frac{\chi}{2}).
\eea
In the Poincare disc coordinates the $AdS_3$ metric is:
 \be
\label{rt3}ds^2=-\left(\frac{1+r^2}{1-r^2}\right)^2dt^2+
\left(\frac{2}{1-r^2}\right)^2\left(dr^2+
r^2 d\phi^2\right),
\ee
where $r<1$.

Let us consider a massless particle with a
lightlike momentum vector pointing along the $x$-direction. Its holonomy
is:

\be\label{uisometry}
u=1+\tan \epsilon \left(
             \begin{array}{cc}
              0  &  2\\
              0 &  0 \\
             \end{array}
              \right),\,\,\,\,\,u^{-1}=1-\tan \epsilon \left(
             \begin{array}{cc}
              0  &  2\\
              0 &  0 \\
             \end{array}
              \right).
\ee

The isometry transformation related to the holonomy has the form

\be
\label{x-star}
{\bf x}\to{\bf x}^*= u^{-1}{\bf x}\,u.\ee

Lightlike particle worldline is the set of fixed points of this isometry, with $r = \tan(t/2)$ and
$\phi=0$. To construct the region that the particle cuts out from
$AdS$ space, one proceeds as following \cite{Matschull:1998rv}. First, we
switch to the ADM-like point of view, so that  $AdS$ can be considered as a Poincar\'e disc evolving in time. Then, one looks for
a pair of some special curves w$_\pm$ on the constant time sections. These curves are
mapped onto each other by the given isometry. Finally, we cut out the
wedge between these curves, and identify the faces according to the
isometry.

Note, that it takes only a finite amount of time for the particle
to travel through the whole space. The particle start position is at $t=-\pi/2$, and the final position is
 at $t=\pi/2$. We consider deformation only for this time interval $-\pi/2<t<\pi/2$ and we expect the space manifold to be a Poincar\'e
disc with a wedge cut out.

 A point $(t,r,-\phi)\in\,{\mbox{w}}_-$ is mapped onto $(t,r,\phi)\in\,{\mbox{w}}_+$ under the isometry action.
The matrices representing these points are
\begin{equation}
 {\bf w}_\pm = \frac{1+r^2}{1-r^2} \, {\bf \Omega}(t)
         + \frac{2\, r}{1-r^2} \, {\bf \Gamma}(\pm\phi).
\end{equation}
Writing the relation
\be
\uu \ww_+=\ww_-\uu\ee
one finds \cite{Matschull:1998rv}, that the faces
w$_+$ and w$_-$ are uniquely determined by the following
equations
\be\label{faces1}
   {\mbox {w}}_\pm:\quad \frac{2\, r}{1+r^2} \, \sin(\epsilon\pm\phi)
          =  \sin t \, \sin \epsilon.
          \ee
The curves w$_+$ and w$_-$
intersect at the
the fixed point of the isometry, lightlike worldline
\be r=\tan(t/2).
\ee
  The spacetime manifold is obtained by cutting out the wedge
behind the particle and identifying the faces of the wedge defined by equations \eqref{faces1}. The resulting spacetime manifold has constant curvature
everywhere, excepting the world line.

\begin{figure}[htbp]
\begin{center}\label{2dwedge}
\includegraphics[width=5cm]{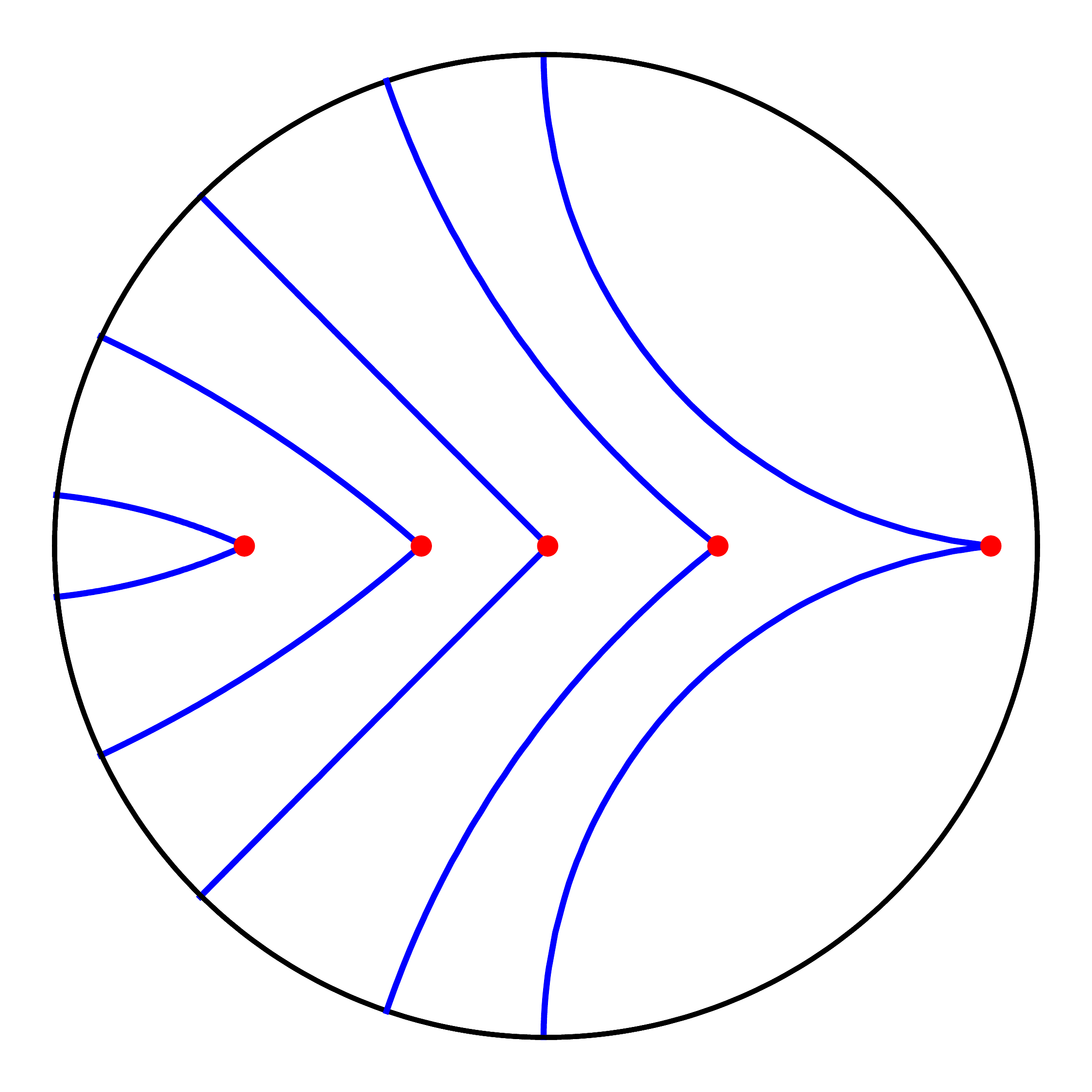}A\,\,\,\,\,\,\,\,\,\,\,\,\,\,\includegraphics[width=5cm]{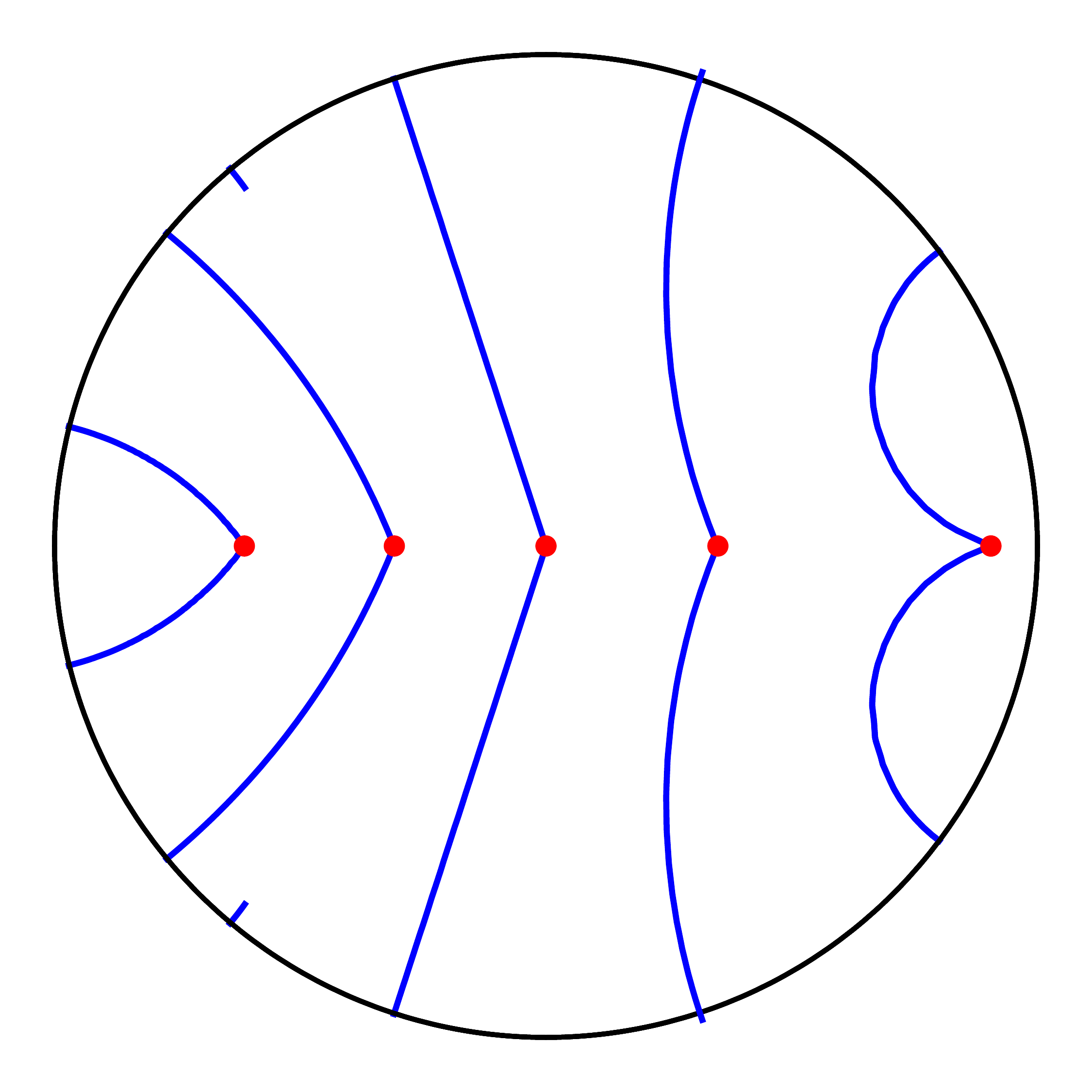}B
\caption{The plot of the wedges $w_\pm$, $\epsilon=\pi/4$ (A) and $\epsilon=0.45\pi$ (B), for different constant time sections. From the left to right(on each plot)  curves correspond to constant time $t$ sections for  $t=-1.1,\,-0.4,\,0,\,0.3,\,1.47$.}
\label{Fig:1moving}
\end{center}
\end{figure}

\begin{figure}[htbp]
\begin{center}\label{3dmoving}
\includegraphics[width=6cm]{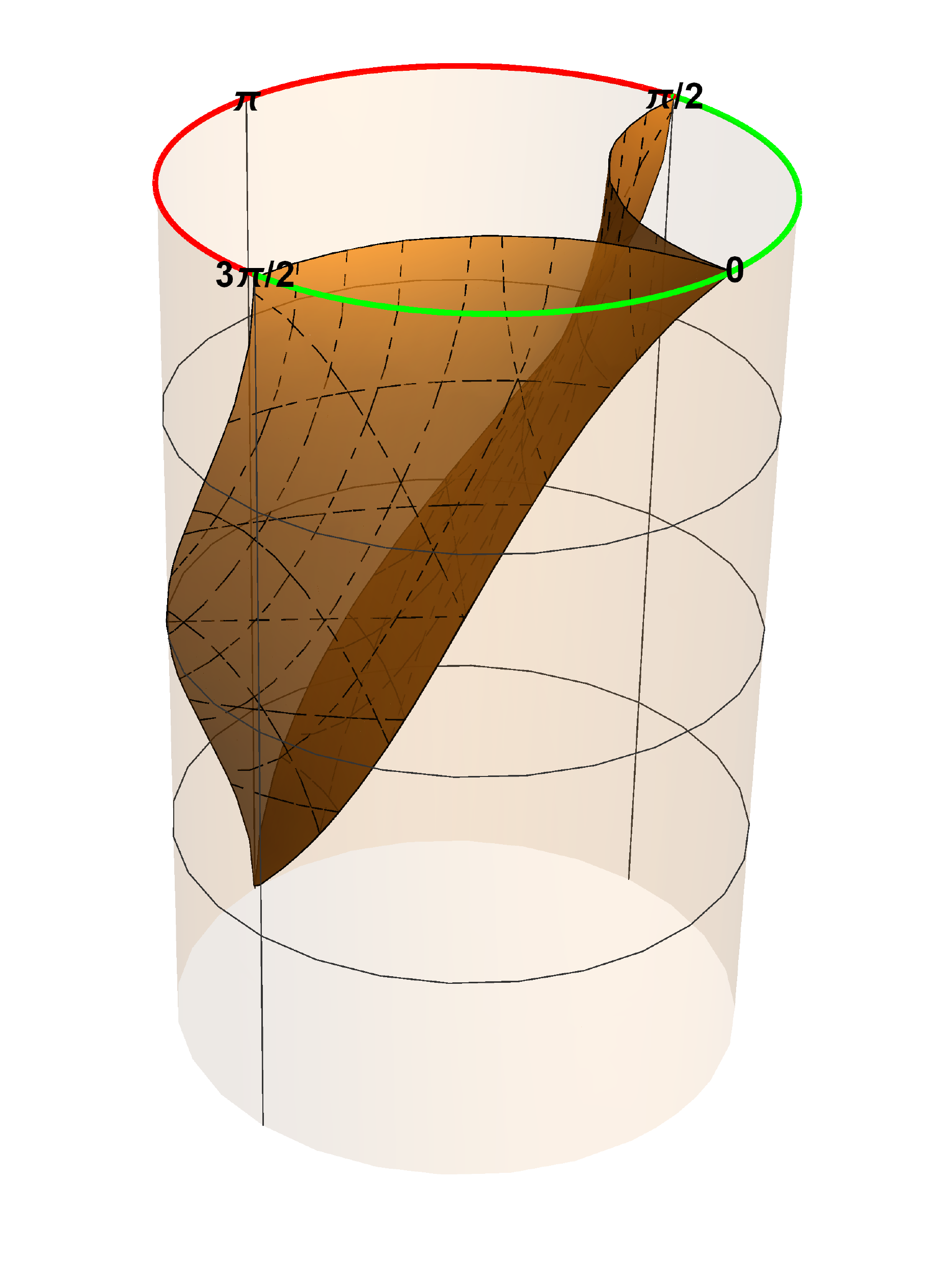}$A$\,\,\,\,\,\,\,\,\,\,\,\,\,\,\includegraphics[width=5.7cm]{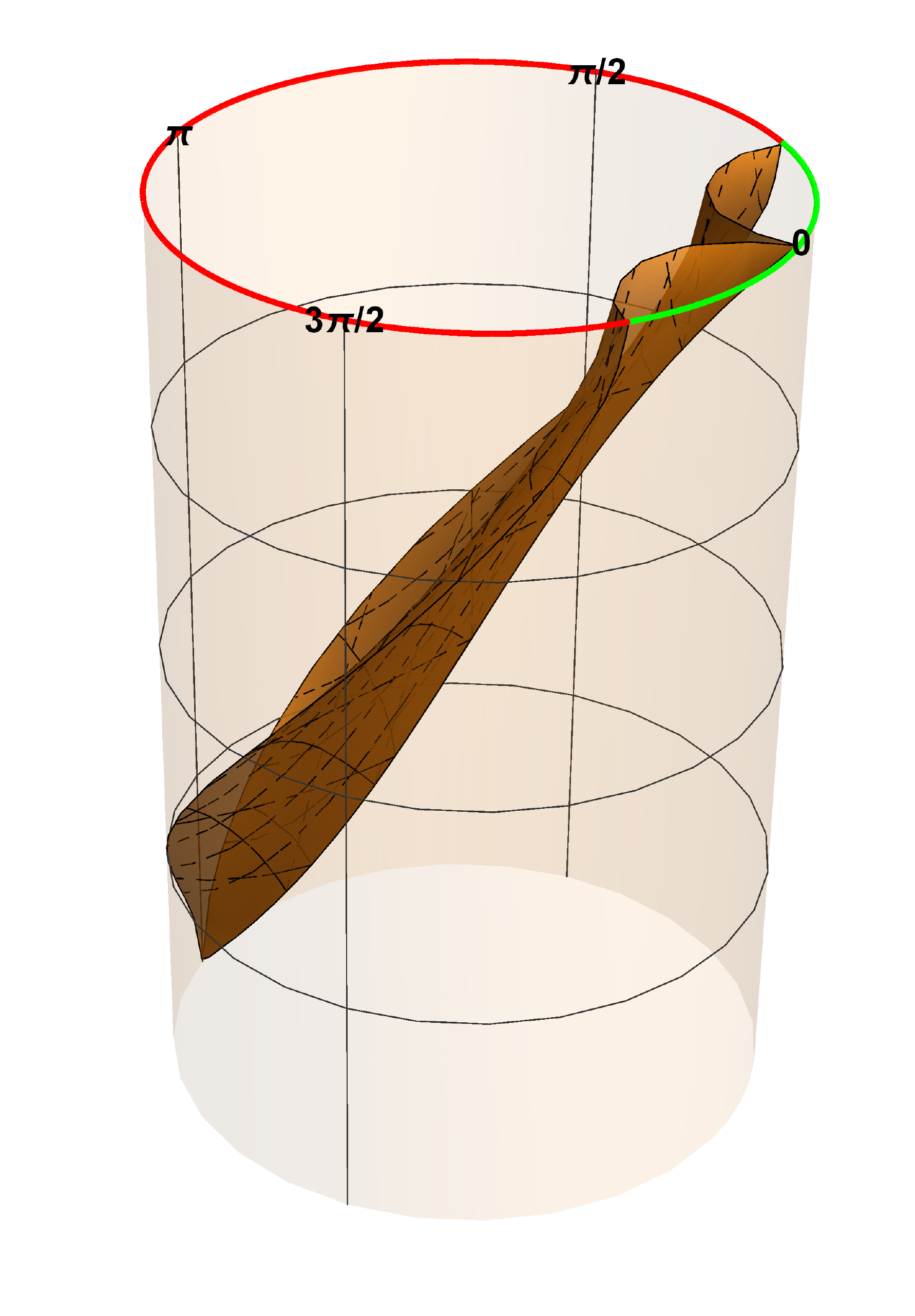}$B$
\caption{The plot of wedge faces that the point particle produces. The green line on the boundary  shows the allowed zone for the final time moment, $t=\pi/2$, the red coloring corresponds to the cut out space. Here we take  $\epsilon=\pi/4$ (A) and $\epsilon=0.45\pi$ (B).}
\label{Fig:3dmoving}
\end{center}
\end{figure}

In Fig.\ref{Fig:1moving} these curves are shown for different constant time sections.
In Fig.\ref{Fig:3dmoving} the wedge to be removed is shown.

In \cite{Matschull:1998rv} the $AdS_3$ space deformed by two ultrarelativistic particles starting from the opposite points
of the $AdS_3$ boundary has been also considered.

As it was mentioned, point sources deform $AdS_3$ in a local way, so for each
particle and wedge faces we can take the result  just mentioned above.
In Fig.\ref{2dwedge-double}. A we plot the process of collision for two ultrarelativistic particles starting with angles $\phi=0$ and $\phi=\pi$ at the moment $t=-\pi/2$, i.e. before $t=0$. The space between faces of the wedges on the left and on the right side are deleted. The picture after collision is presented in Fig.\ref{2dwedge-double}.B.

\begin{figure}[htbp]
\begin{center}\label{2dwedge-double}
\includegraphics[width=5cm]{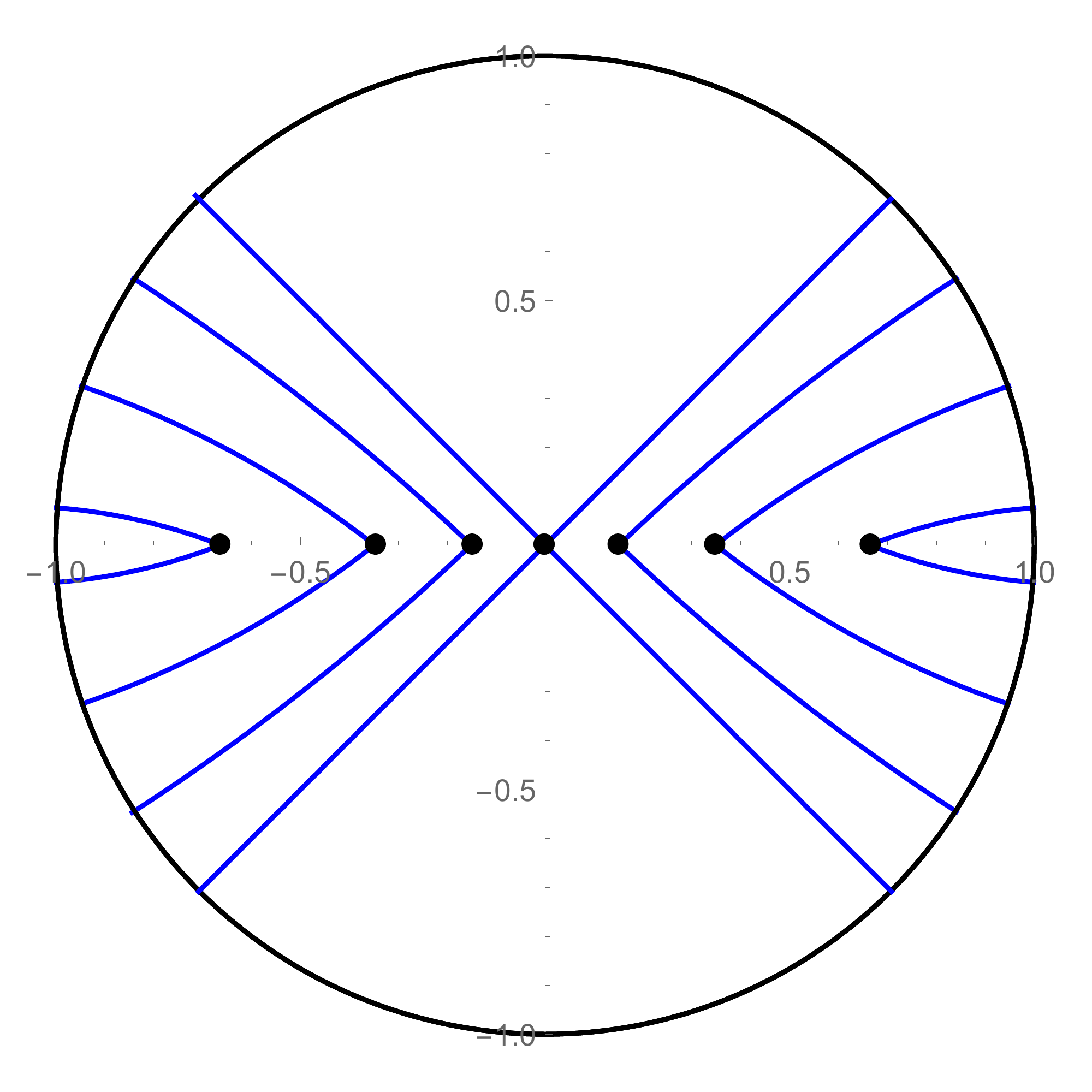}A\,\,\,\,\,\,\,\,\,\,\,\,\,\,\includegraphics[width=5cm]{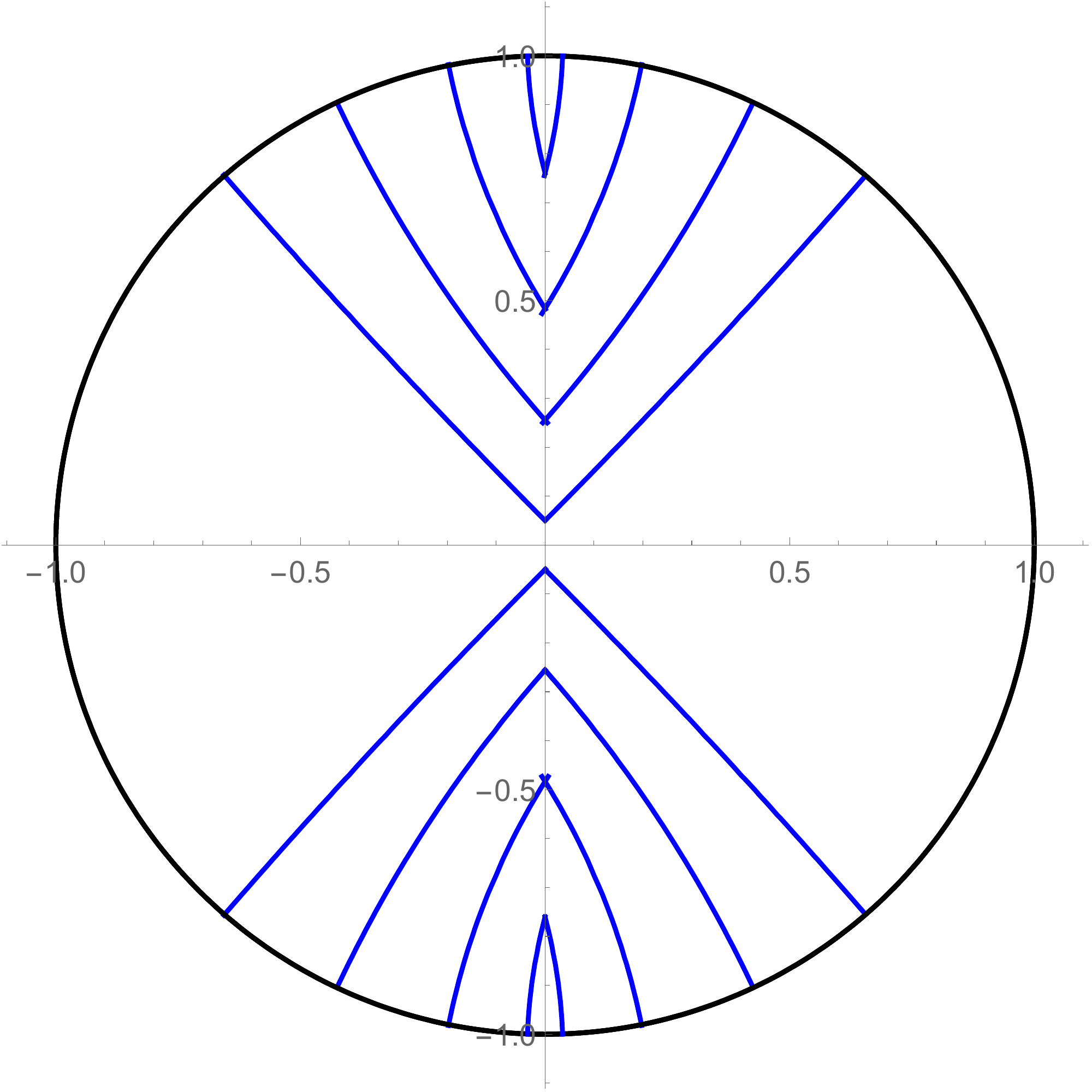}B
\caption{The plot of two colliding massless particles picture for $\epsilon=\pi/4$, before collision (A) and after collision (B) for different times $t$. On the left plot $t=-1.17,\,-0.67,\,-0.3,\,0.$. In  the right plot $t=0.1,\,0.5,\,0.9,\,1.3$.}
\label{2dwedge-double}
\end{center}
\end{figure}

\section{Correlators on the boundary of $AdS_3$ deformed by moving defects in the bulk}

\subsection{One  ultra-relativistic defect}

Let us consider the two-point correlator on the boundary of $AdS_3$ deformed by one massless particle\footnote{Here we mean a deformation of the universal two-point correlator, given by eq.(2.76) in \cite{AAT}. It is related with the Wightman, retarded and causal correlators by well-known formulae, see Sec.2.3.4. in \cite{AAT}.}

 The two-point correlator on the boundary of $AdS_3$ deformed by one massless particle is:

 \bea\label{cor*}
G_{\epsilon,\Delta}(\phi_a,t_a,\phi_b,t_b)&=&\left(\frac{1}{2\left|\cos(t_a-t_b)-\cos(\phi_a-\phi_b)\right|}\right)^\Delta\,\Theta_{ncr}(t_a,\phi_a;t_b,\phi_b;\epsilon)\\\nn &+&
\left(\frac{1}{2\left|\cos(t_{a^\#}-t_b)-\cos(\phi_{a^\#}-\phi_b)\right|}\,C_{a^\#}^{-1/2}\right)^\Delta
\,\Theta_{cr}(t_{a^\#},\phi_{a^\#};t_b,\phi_b;\epsilon).
 \eea
where

\bea
C_{b^*}&=&\left( \left(2 \tan ^2(\epsilon )+1\right) \sin (t_b)-2  \tan (\epsilon ) \sec (\epsilon ) \sin (\epsilon -\phi_b )\right)^2+\cos ^2(t_b)\\
C_{b^\#}&=&\left( \left(2 \tan ^2(\epsilon )+1\right) \sin (t_b)-2  \tan (\epsilon ) \sec (\epsilon ) \sin (\epsilon +\phi_b )\right)^2+\cos ^2(t_b)
\eea
are the renormalization factors (see \cite{AAT} for more details).
The function $\Theta_{ncr}$ is defined as follows:
\begin{itemize}
\item $\Theta_{ncr}(t_a,\phi_a;t_b,\phi_b;\epsilon)=1$ if geodesic connecting points $(\phi_a,t_a)$ and $(\phi_b,t_b)$ does not cross the wedge at some time;
\item $\Theta_{ncr}(t_a,\phi_a;t_b,\phi_b;\epsilon)=0$ if geodesic connecting points $(\phi_a,t_a)$ and $(\phi_b,t_b)$ crosses the wedge at some time.
\end{itemize}
$\Theta_{cr}$ is defined as follows:
\begin{itemize}
\item$ \Theta_{cr}(t_a,\phi_a;t_b,\phi_b)=1$ if geodesic crosses  the nearest from the point $b$ face of the wedge at any time;
\item $\Theta_{cr}(t_a,\phi_a;t_b,\phi_b)=0$ if geodesic  does not cross the nearest face of the wedge from the point $b$.
\end{itemize}

Calculating supports of functions $\Theta_{cr}$ and $\Theta_{ncr}$ numerically we find which geodesics  contribute to the correlator. Generally speaking, there are  three different possibilities for geodesic configurations connecting two points on the boundary:
\begin{itemize}
\item there is a geodesic connecting $a$ and $b$ that does not cross the wedges (we call this geodesic the basic, or the
non-winding one) and there is no winding geodesic connecting $a$ and $b$ (i.e. there is no geodesic connecting $a$ and $b$
and crossing the wedge)
\item there is a winding geodesic connecting $a$ and $b$ and there is  no one geodesic connecting $a$ and $b$
\item there is  a basic geodesic connecting $a$ and $b$ and simultaneously there is a winding geodesic connecting $a$ and $b$.
\end{itemize}
In Fig.\ref{geodconf} and Fig.\ref{geodconf2} we plot the different cases of  geodesic configurations for different values of the parameter $\epsilon$. If the basic geodesic does not cross the wedge  it contributes to the propagator, and conversely, if the geodesics connecting images does not cross the wedge, it does not act as a winding geodesic.

\begin{figure}[h]\label{geodconf}
   \centering
    \begin{picture}(90,180)
\put(-170,5){\includegraphics[width=4cm]{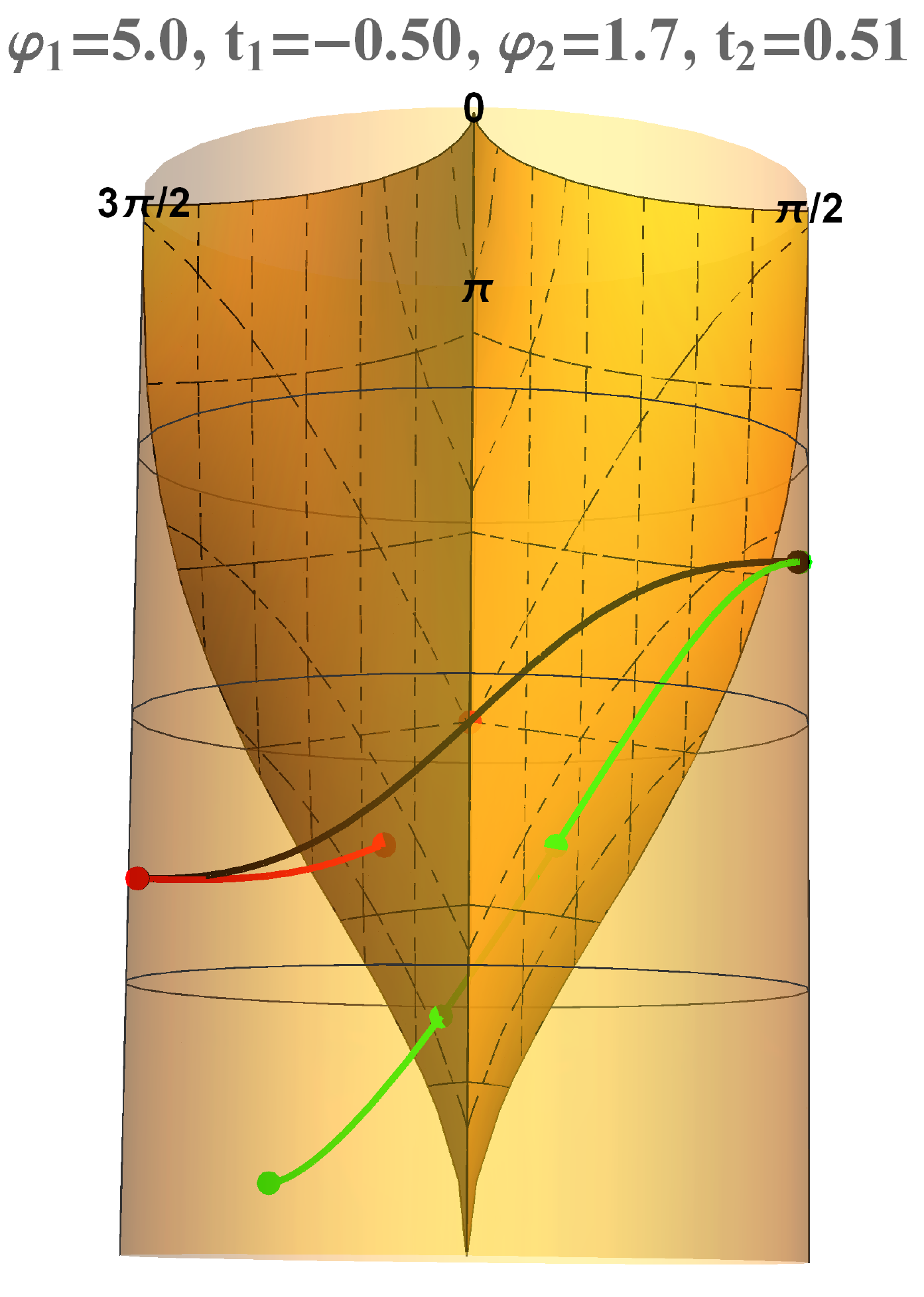}}
\put(-35,-13){\includegraphics[width=4.6cm]{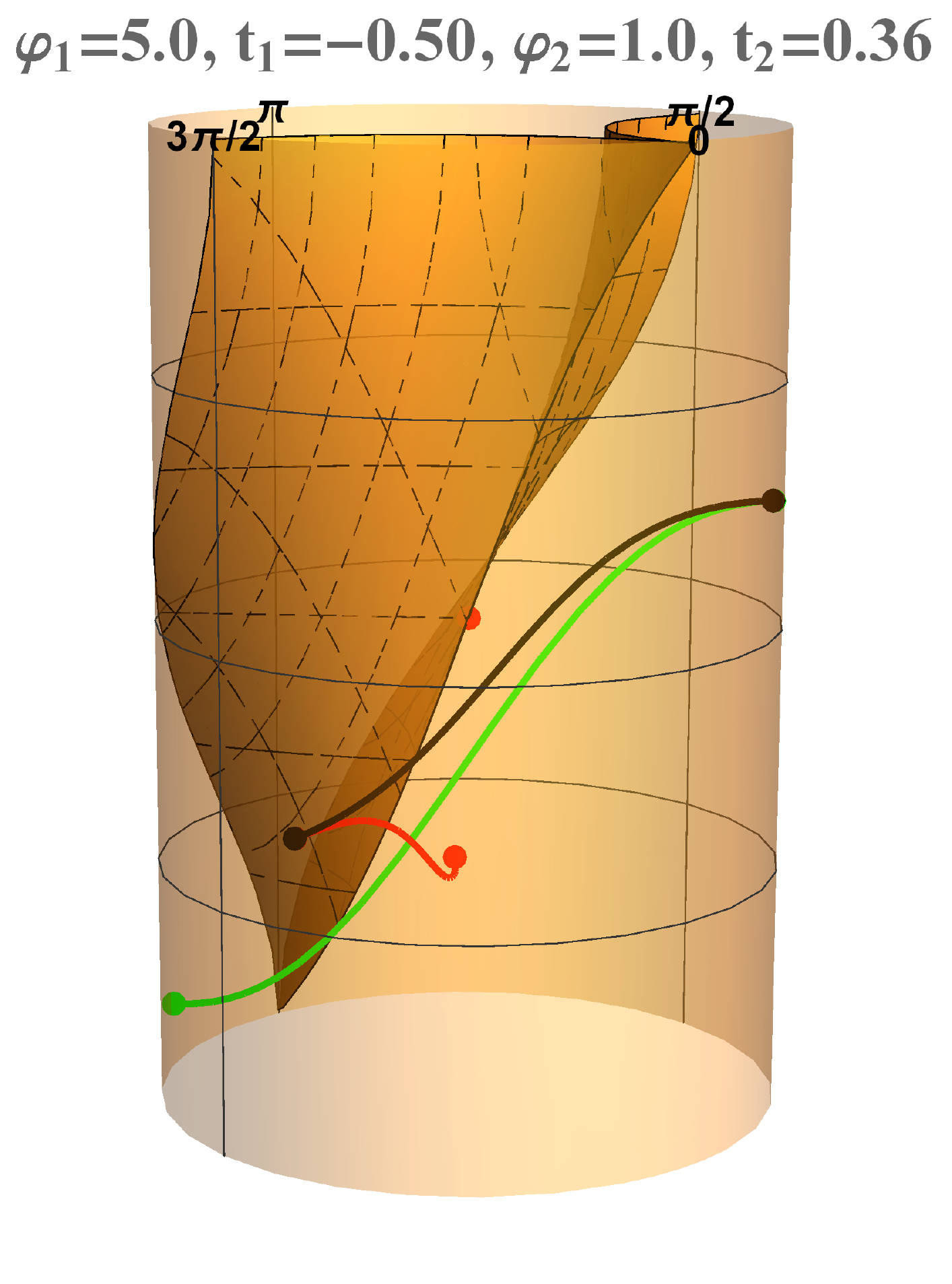}}
\put(125,2){\includegraphics[width=4.6cm]{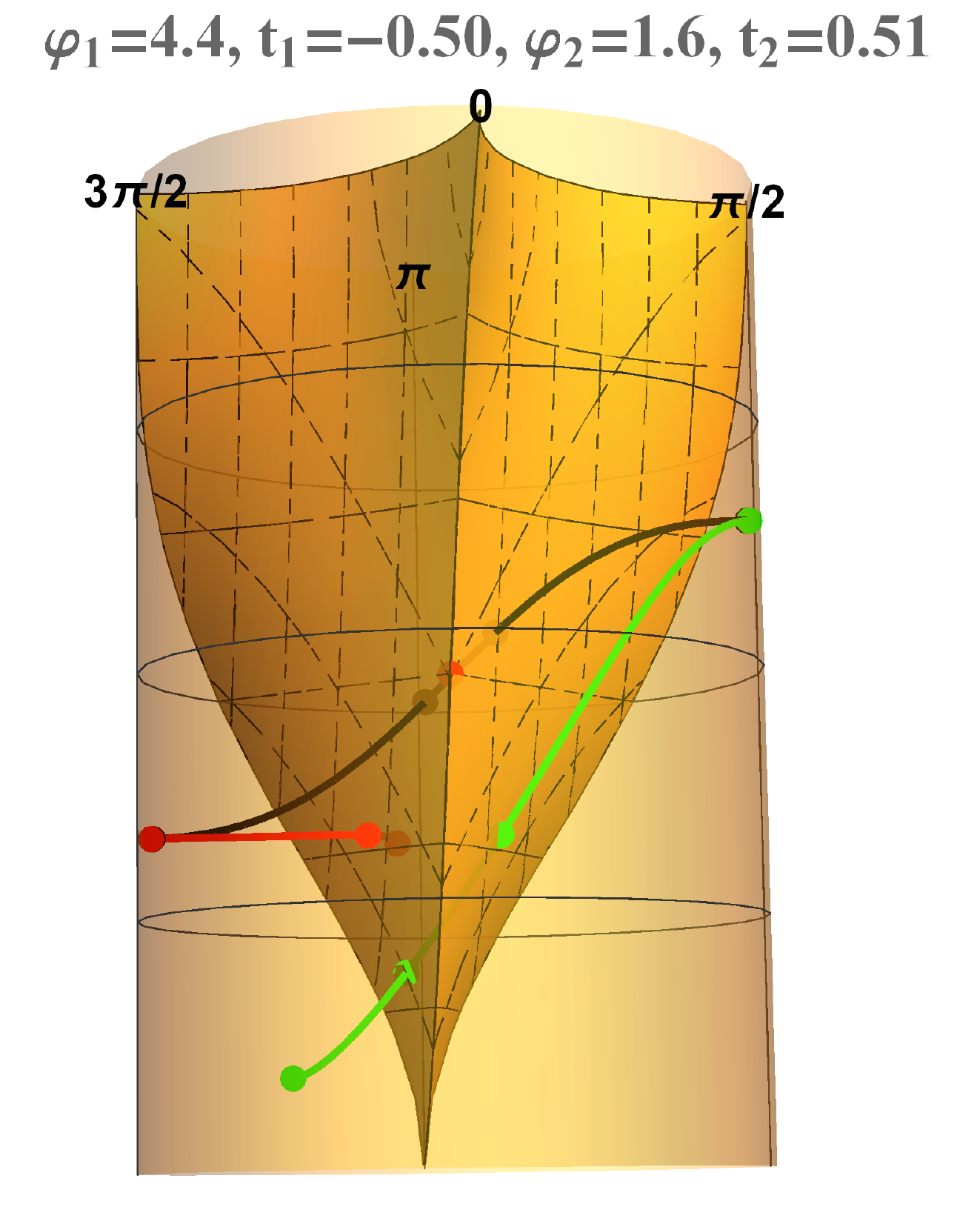}}
\put(-60,-5){A}
\put(90,-5){B}
\put(235,-5){C}
\end{picture}\\$\,$\\
    \caption{ The black curves on plots A, B and C are the basic geodesics, the red and green ones are winding geodesics. Boundary points ($\phi_a,t_a$) and ($\phi_b,t_b$) correspond to the black geodesic endpoints.   $\epsilon = \frac{\pi}{4}.$
   }
  \label{geodconf}
\end{figure}
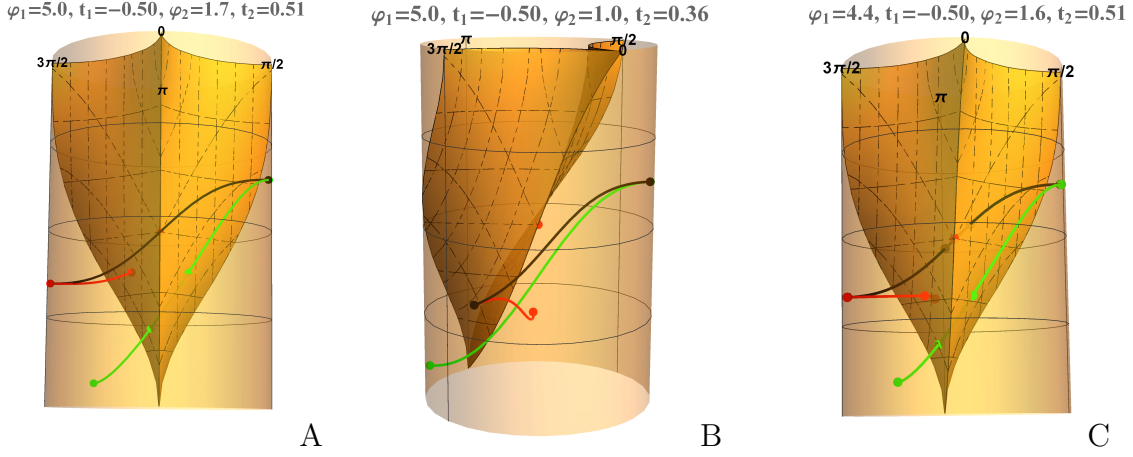

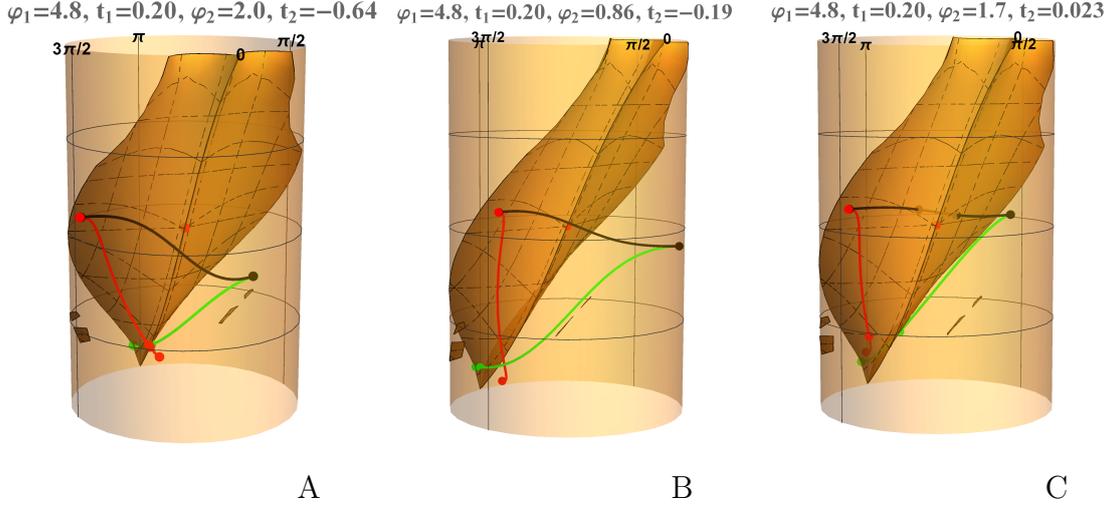
\begin{figure}[h]\label{geodconf2}
  \centering
   \begin{picture}(90,210)
\put(-160,5){\includegraphics[width=5cm]{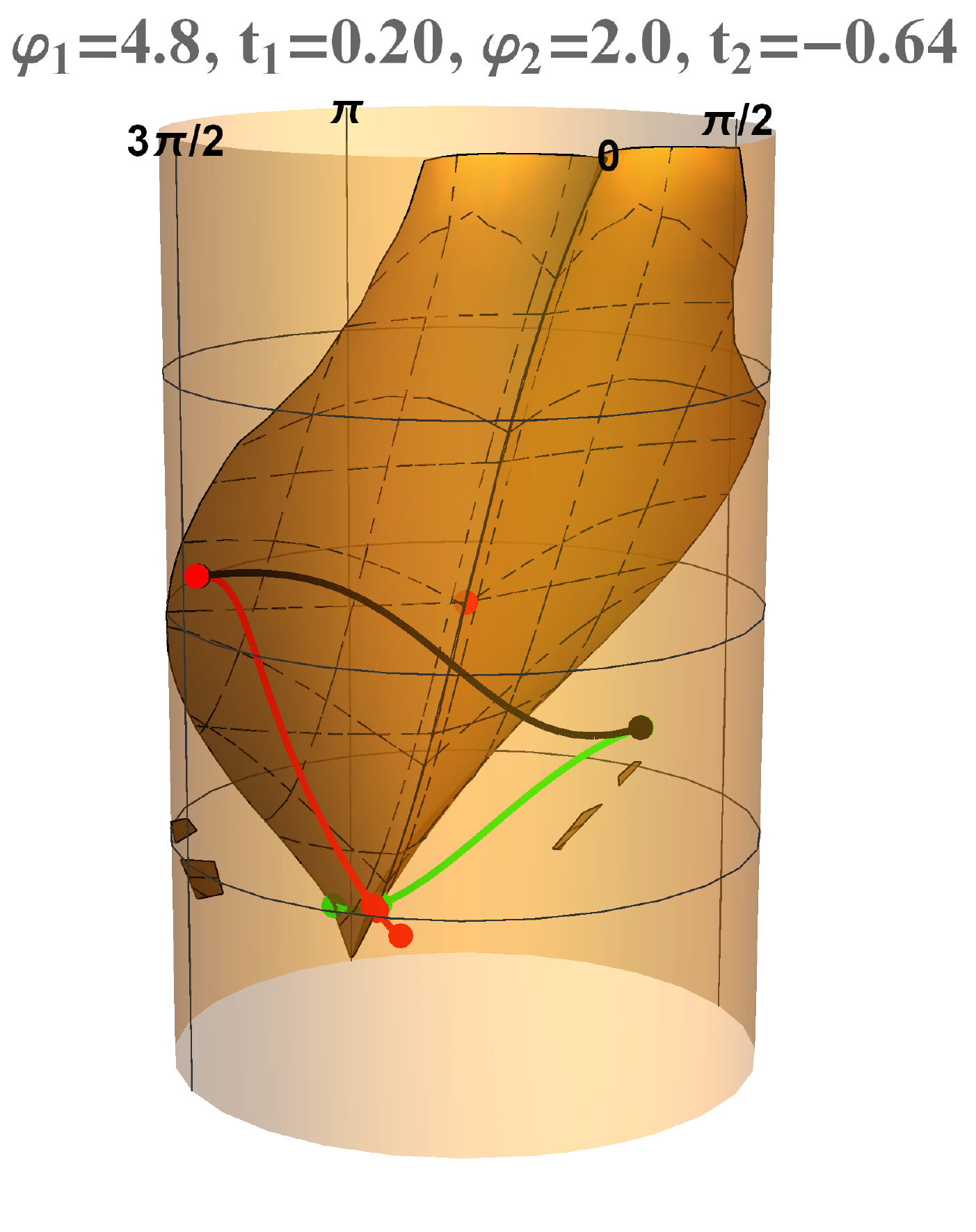}}
\put(-15,0){\includegraphics[width=4.62cm]{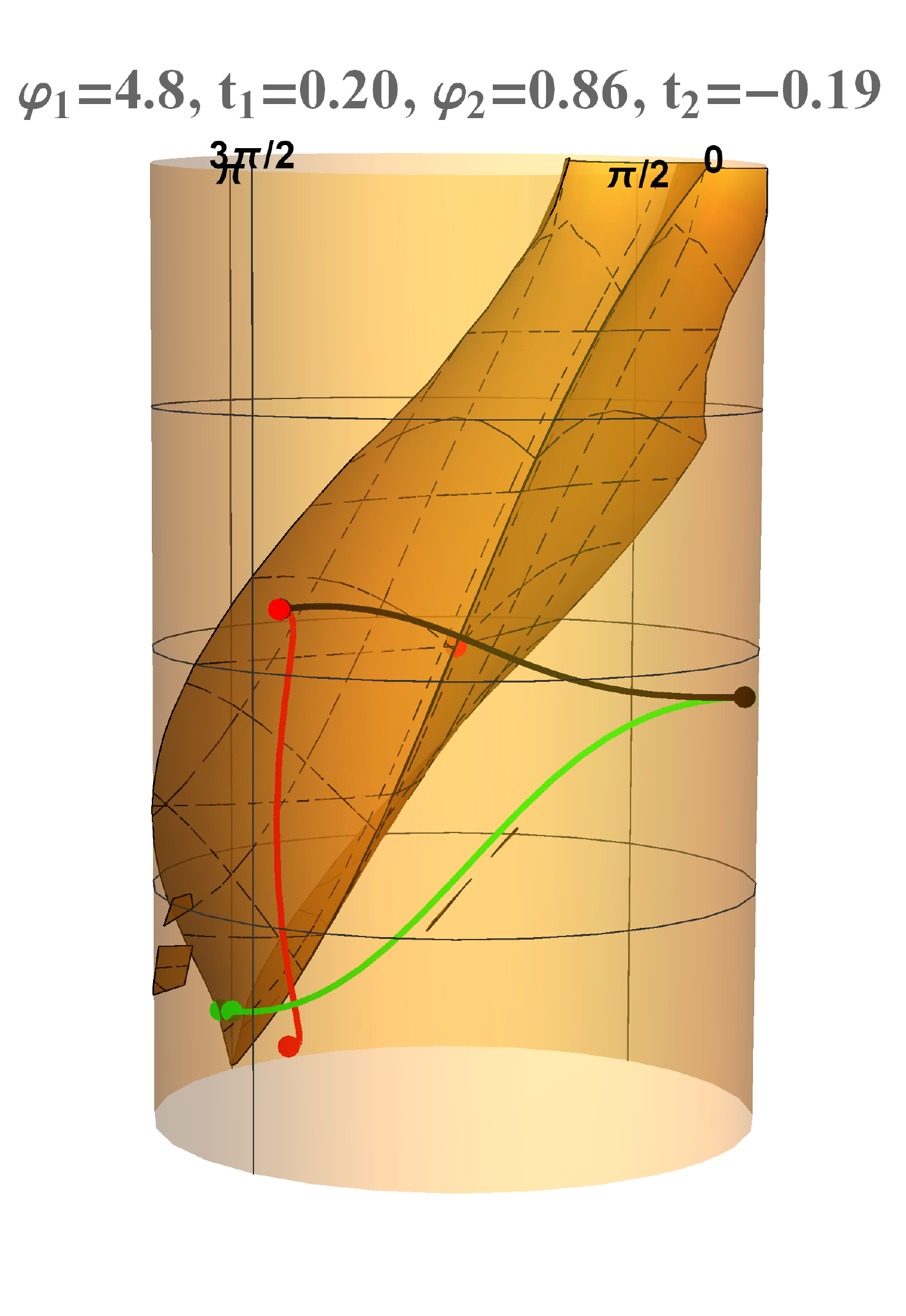}}
\put(125,0){\includegraphics[width=4.6cm]{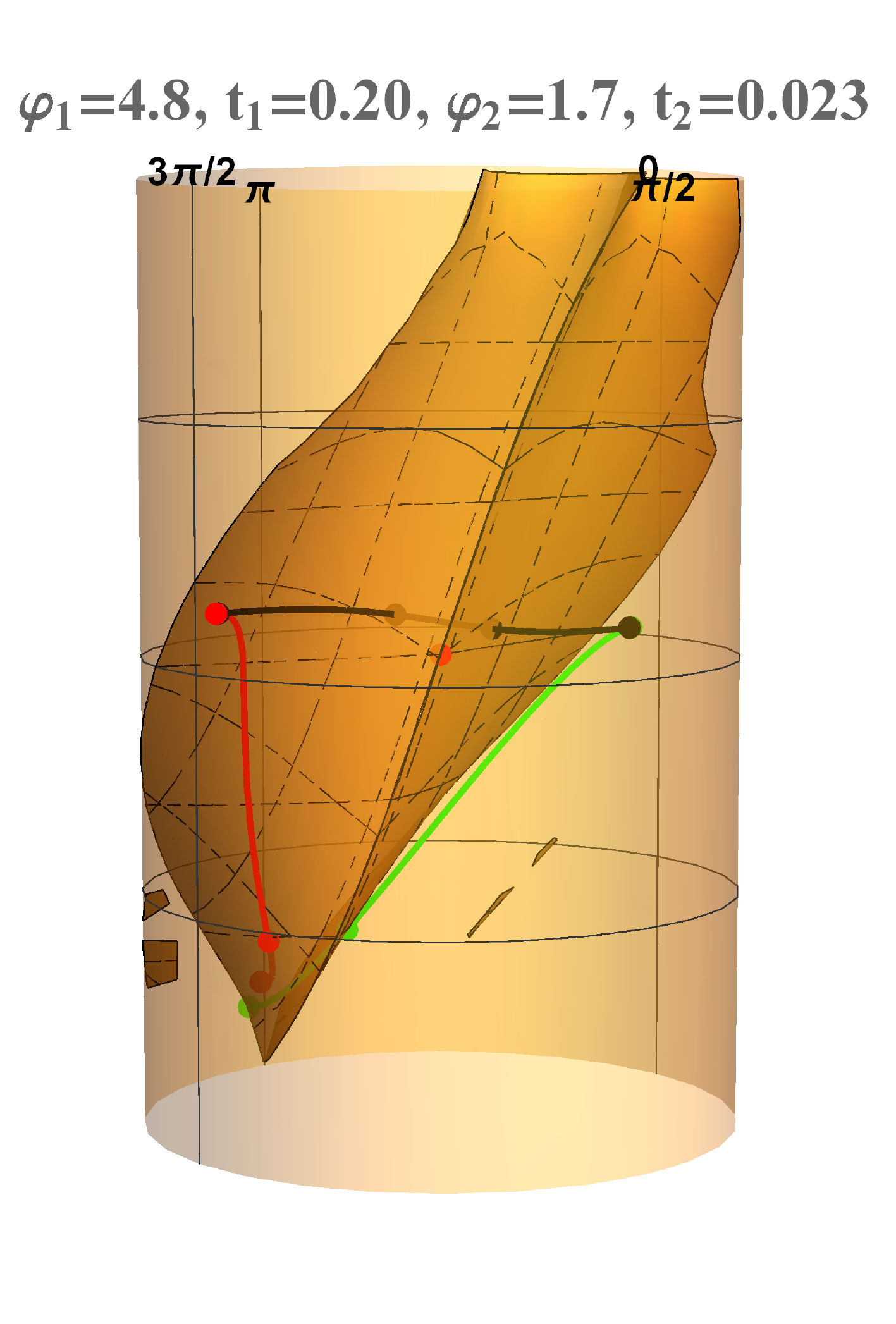}}
\put(-50,-5){A}
\put(90,-5){B}
\put(230,-5){C}
\end{picture}\\$\,$
      \caption{ The black curves on plots A, B and C are the basic geodesic, the red and green are winding ones. Boundary points ($\phi_a,t_a$) and ($\phi_b,t_b$) correspond to the black curve endpoints. $\epsilon =0.4\pi$ .
   }
  \label{geodconf2}
\end{figure}

In Fig.\ref{geodconf}.A the black curve does not  intersect the wedges and winding geodesics  (green and red lines) intersect  the wedges. In this case both contribute in \eqref{cor*}. In Fig.\ref{geodconf}.B the black curve is noncrossing and red and green curves do not cross the wedge faces too. In this case  only the non-winding geodesic (the black curve) contributes. In Fig.\ref{geodconf}.C  the black curve intersects the wedge, geodesics  relating corresponding image points also intersect the wedge, so only the winding geodesic contributes. The same picture takes place for geodesics presented in Fig.\ref{geodconf2}. A, B and C.

\begin{figure}[h]
$$\,$$
 \centering
\includegraphics[width=6cm]{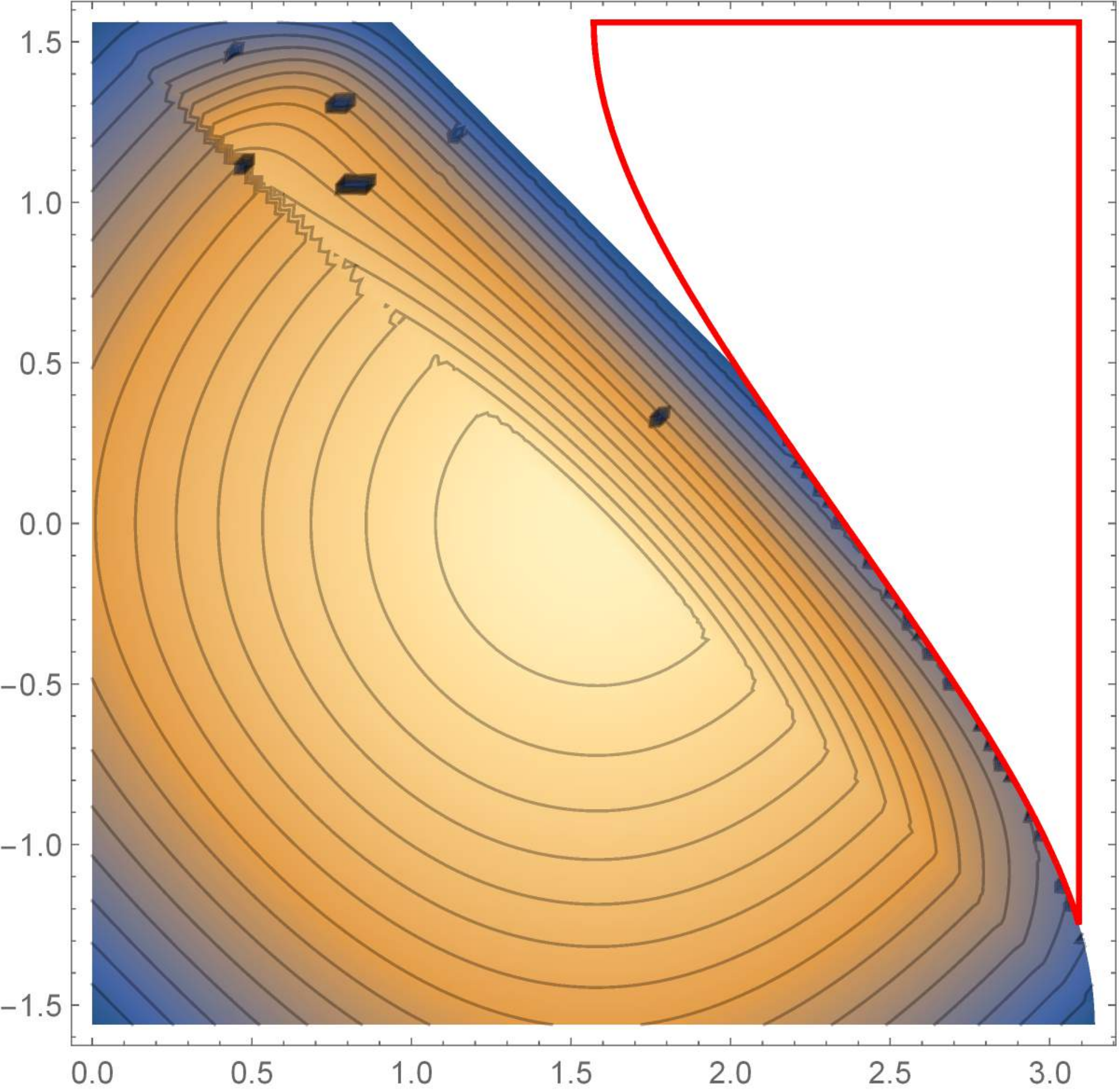}A.
 $\,\,\,\,\,$
 \includegraphics[width=6cm]{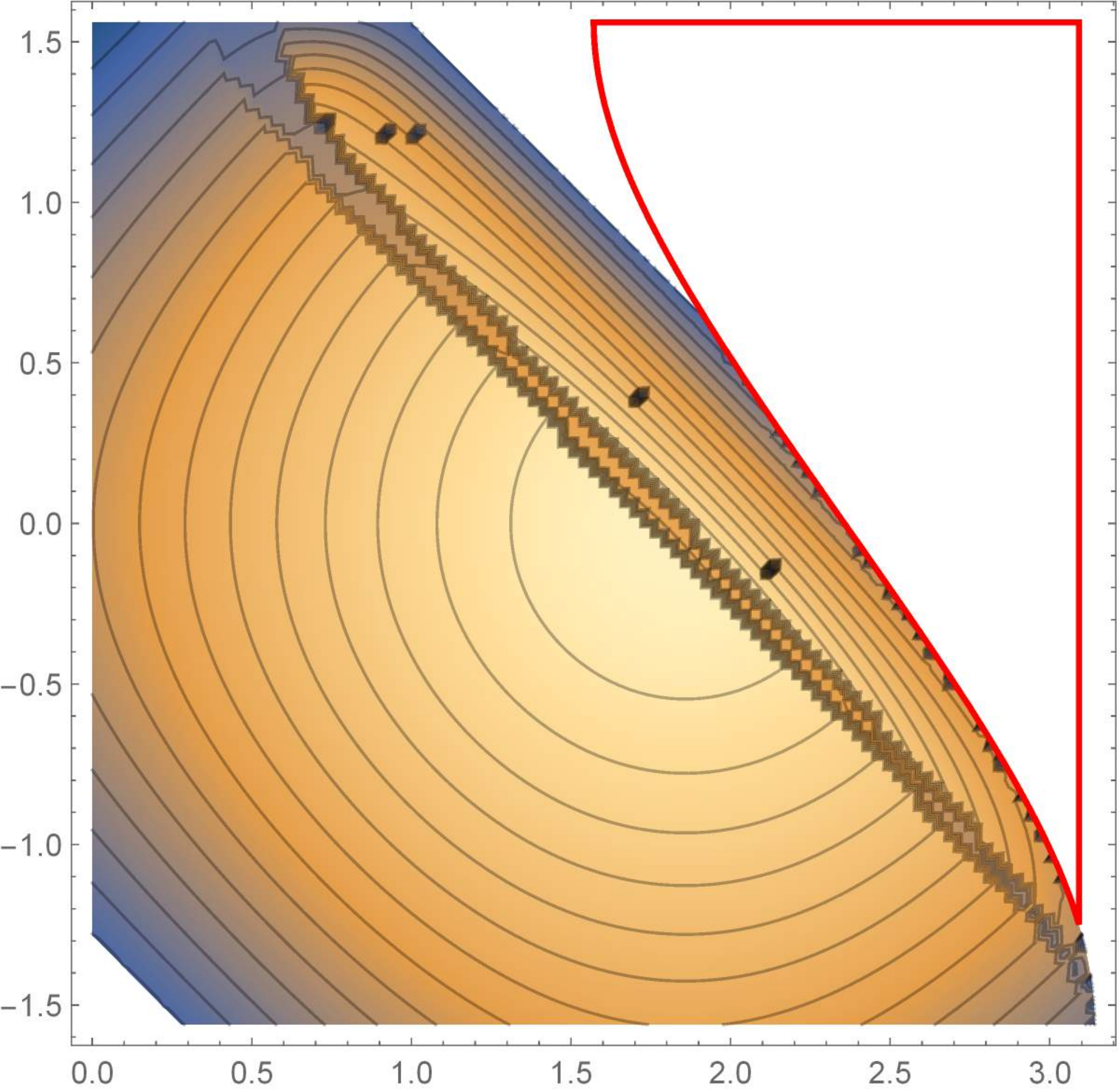}B.
 \caption{Density plot of the function \eqref{G1}. The angular coordinate corresponds to the x-axis and time coodinate to the y-axis. The point particle parameter value $\epsilon =\frac{\pi}{4}$. The red thick curve corresponds to the boundary of the living space.
 A. Fixed angle $\phi_1=\frac{3\pi}{2}$ and time $t_1=0$, conformal dimension value is $\Delta=1$.
  The thick black line near the diagonal  corresponds to the discontinuity in the correlator due to the presence of different contributions.
 B. Fixed angle $\phi_1=5$ and time $t_1=0$, conformal dimension value is $\Delta=1$. Two thick black lines near the diagonal correspond
  to the discontinuity in the correlator due to different contributions presence. }
  \label{2dim3pi2}
\end{figure}

\begin{figure}[h]   \centering
$$\,$$\\
\includegraphics[width=6cm]{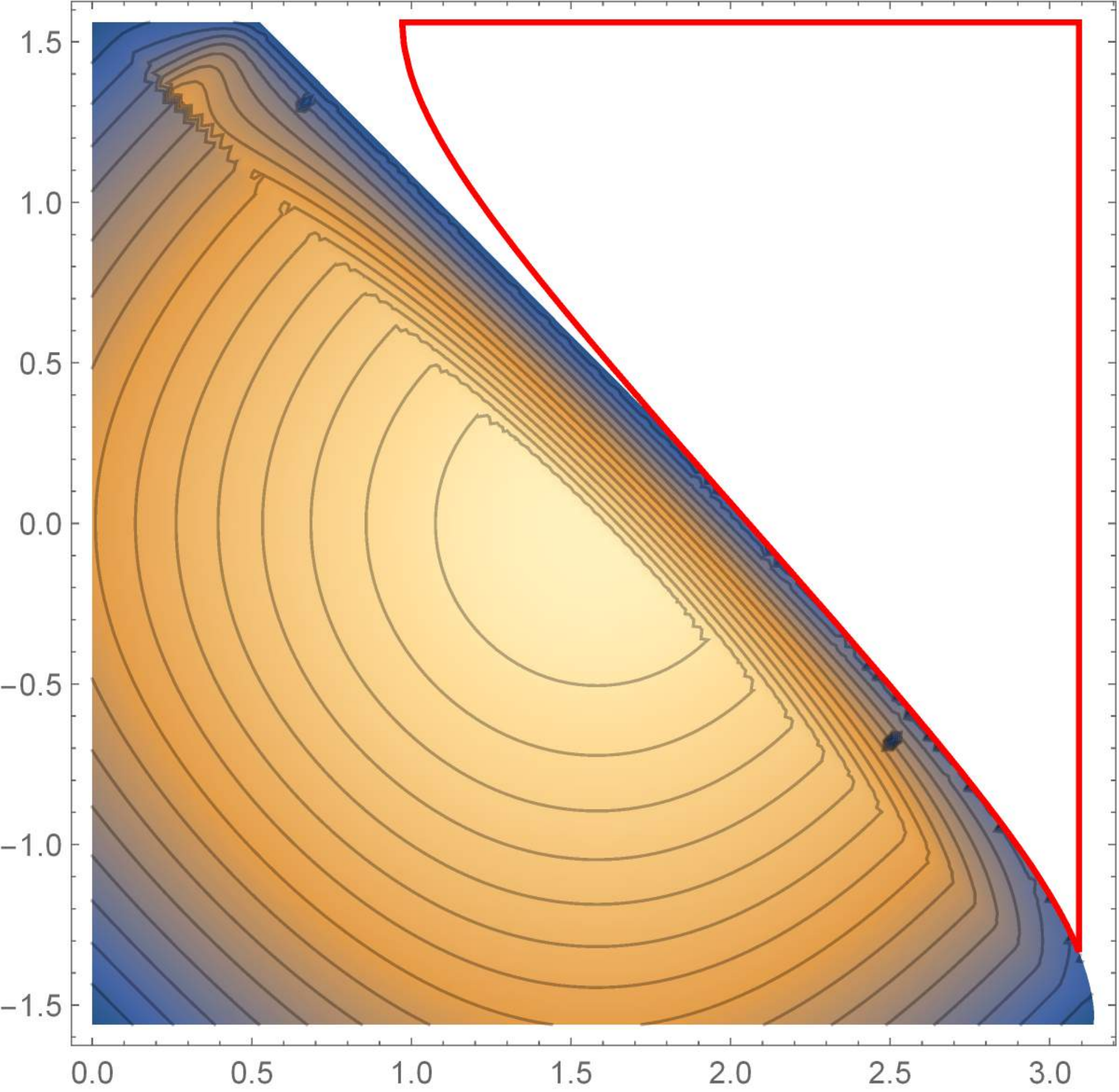}A.
 $\,\,\,\,\,$
  \includegraphics[width=6cm]{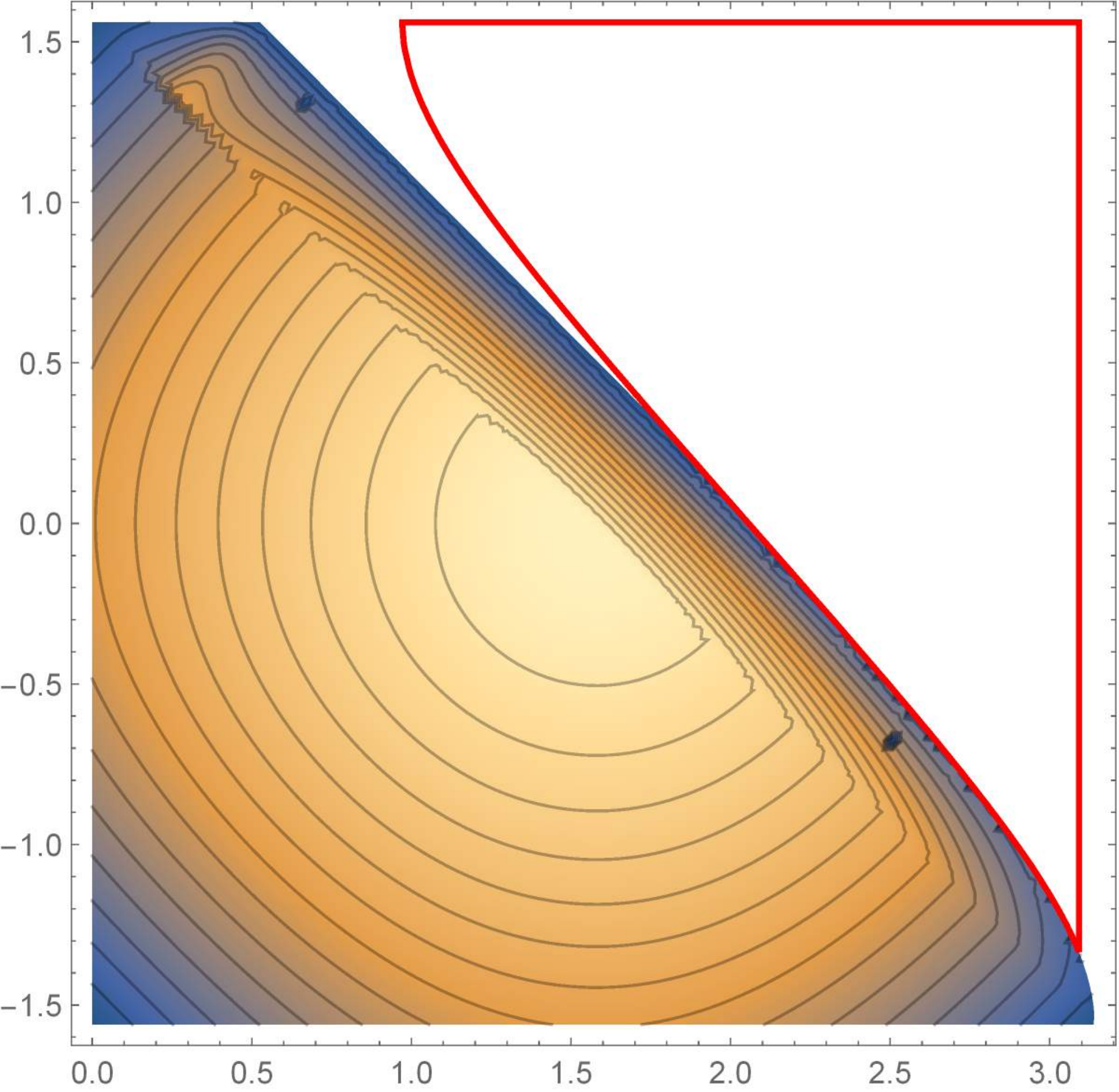}B.
  \\
  \caption{Density plot of the functions \eqref{G1}. The point particle parameter value $\epsilon=1$.  The angular coordinate corresponds to the x-axis and time coordinate to the y-axis. The red thick curve corresponds to the boundary of the living space. The conformal dimension is $\Delta=1$ for  both plots A. and B.
  A. Fixed angle $\phi_1=\frac{3\pi}{2}$ and time $t_1=0$.
   The thick black line near the diagonal  corresponds to the discontinuity in the correlator due to the presence of different contributions.
   B. Fixed angle $\phi_1=5$ and time $t_1=0$. Two thick black lines near the diagonal correspond
   to the discontinuity in the correlator due to different contributions presence}\label{2dim3pi2s}
\end{figure}

Let us see the influence of moving particle on the two-point function $G_{\epsilon,\Delta}(\phi_1,t_1,\phi_2,t_2)$.
Let us take  one of the points, namely $(\phi_1,t_1)$, to be fixed. Suppose, that  $\phi_1$ and $\phi_2$ are located on the opposite halves of the $AdS_3$ boundary with respect to the lightlike worldline, i.e. $\phi_1 \in(\pi,2\pi)$, $\phi_2 \in (0,\pi)$, and time $t_1,t_2 \in (-\frac{\pi}{2},\frac{\pi}{2})$, i.e. we study the quantity:

\be\label{G1}
\mathcal{G}_{\epsilon,\Delta,\phi_1,t_1}\left(\phi,t  \right)=G^{-1}_{\epsilon,\Delta}(\phi_1,t_1,\phi,t).
\ee
 In Fig.\ref{2dim3pi2} and Fig.\ref{2dim3pi2s} we plot $\mathcal{G}_{\epsilon,\Delta,\phi_1,t_1}\left(\phi,t  \right)$ for different $\epsilon$, $t_1$ and $\phi_1$.
 From  Fig.\ref{2dim3pi2} and Fig.\ref{2dim3pi2s} we see, that  near the edge of the living space
 there is a nontrivial change in the correlator. We call this zone a pulse. Pulse propagates from the  edges of the defect and its size is changed with time.
In Fig.\ref{2dim3pi2}.A we can see, that the discontinuity is formed at the moment $t=-\frac{\pi}{2}$,  near the point on the boundary, wherefrom the
 massless particle starts, then this discontinuity propagates with nearly constant speed. On the left side from the discontinuity the correlator is not changed and here the conformal symmetry is not broken.
In Fig.\ref{2dim3pi2}.B we see how additional discontinuity appears, due to mixing of different
geodesic contributions and we get a resonance.

\newpage
\subsection{Two colliding ultra relativistic particles.}

 Let us consider  two colliding point particles in the $AdS_3$. It includes more possibilities for geodesics to wind on the faces of the defect and
the picture is more involved for this case.

\begin{figure}[h]
    \centering
        \includegraphics[width=8cm]{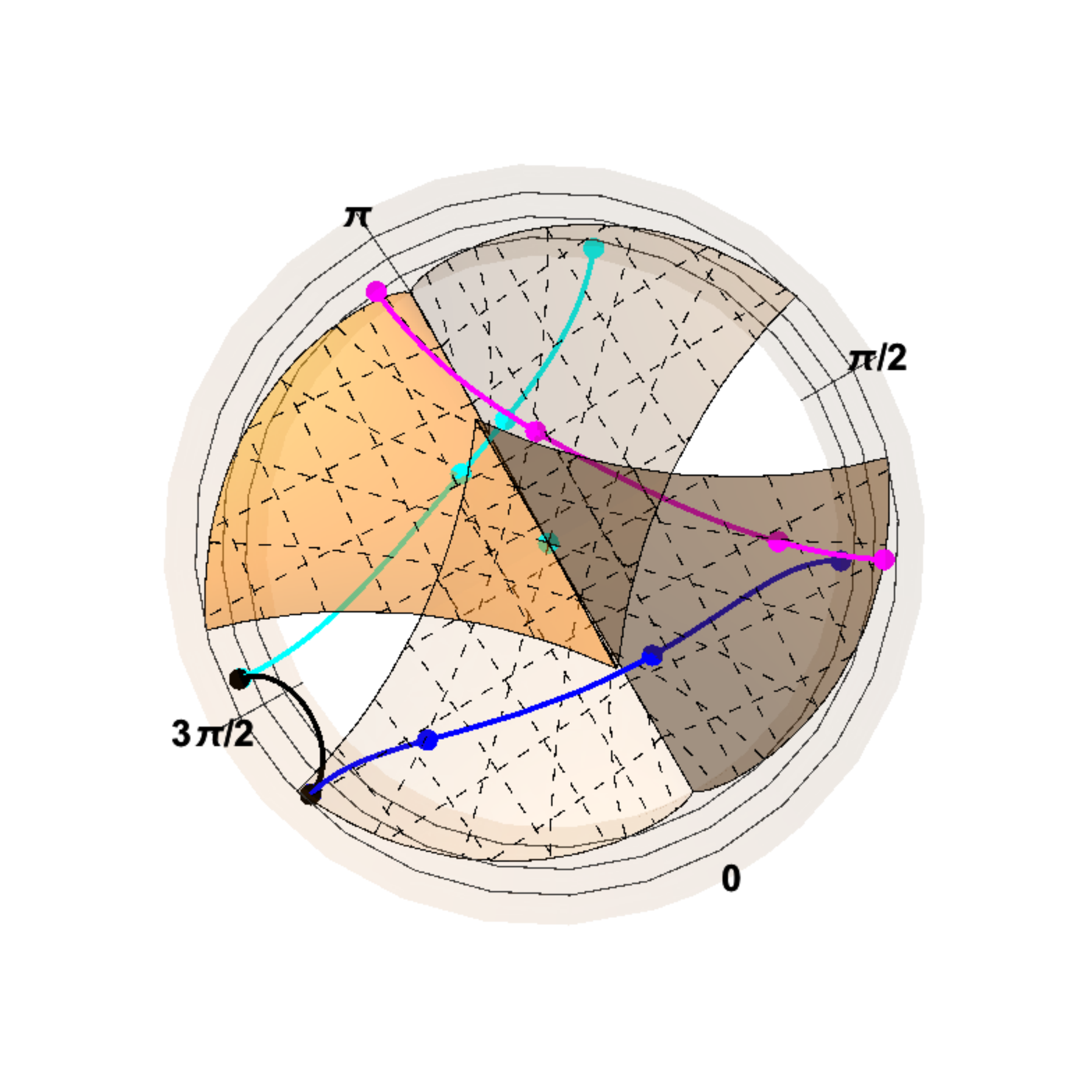}
            \caption{ Illustration of one of the additional contributions to the correlator, when points on the boundary (the black curve endpoints) to be taken on one side of the boundary. They wind through both wedges wedges.   Black curve is the basic geodesic, the magenta geodesic is an intermediate, and it is the result of application of the isometry induced by left wedge to green geodesic. The blue geodesic is the result of application of thee isometry induced by the right wedge to  magenta geodesic. Here we take $\epsilon=\pi/4$ for both particles.}
    \label{triple-w}
\end{figure}

There are two different cases:

\begin{itemize}
\item The first one is when we take two points on the opposite sides of cylinder, symmetrically with respect to the line of collision, i.e. we consider correlator $G(\phi_1,t_1,\phi_2,t_2)$, where $\pi<\phi_1<2\pi$ and $0<\phi_2<\pi$. Time is taken to be $-\pi/2<t_1,t_2<\pi/2$. In this case we have the contribution coming from two types of geodesics: the basic geodesic and the geodesics passing once through each defect (see Fig. \ref{double-w})
\item The second one is when we take both points on one side of the boundary, i.e. $\pi<\phi_1<2\pi$ and $\pi<\phi_2<2\pi$. In this case the two-point correlator obtains two contributions, one from the basic geodesic, and one from the geodesic that winds two times. First it passes through the lower face of the left wedge, then through the upper face of the right wedge. The schematic illustration of this case is presented in Fig. \ref{triple-w}A.
\end{itemize}

In this paper we consider only the first case from the above list. It corresponds to the long-range effects in dual theory.
Let us consider the first situation from the above list.
The universal correlator for the case when points are on the opposite sides of $AdS_3$ is:

\bea \label{corr-opposite}
&\,& G(\phi_a,t_a,\phi_b,t_b)_{\epsilon_1,\epsilon_2,\Delta}\\\nn
&=&\left(\frac{1}{2\left|\cos(t_a-t_b)-\cos(\phi_a-\phi_b)\right|}\right)^\Delta\,
\Theta_{ncr}^{all}(a,b;\epsilon_1)\cdot\Theta_{ncr}^{all}(a;b;\epsilon_2)\\\nn
&+&
\left(\frac{1}{2\left|\cos(t_{a^{\#_1}}-t_b)-\cos(\phi_{a^{\#_1}}-\phi_b)\right|}\,C_{a^{\#_1}}^{-1/2}\right)^\Delta
\Theta_{cr}^{ll}(a^{{\#}_1},b;\epsilon_1)
\cdot \Theta_{ncr}^{lu}(a,b^{*_1};\epsilon_2)
\\\nn &+&
\left(\frac{1}{2\left|\cos(t_a-t_{b^{\#_2}})-\cos(\phi_a-\phi_{b^{\#_2}})\right|}\,C_{a^{\#_2}}^{-1/2}\right)^\Delta
\,\Theta_{cr}^{rl}(a^{\#_2},b;\epsilon_2)\Theta_{ncr}^{ru}(a;b^{*_2};\epsilon_1)
 \\\,\nn
\eea
The term in the first line corresponds to the basic geodesic contribution, the second line corresponds to geodesic winding through the left wedge and the third line corresponds to the contribution from the geodesic winding through the right wedge.
Here we use different functions $\Theta$ defined as following.

 Function $\Theta_{ncr}^{\text{all}}(a,b;\epsilon)=1$  if the geodesic connecting two points $a$ and $b$ does not cross any wedge and $\Theta_{ncr}^{all}(a,b;\epsilon)=0$ otherwise.

  Function $\Theta_{cr}^{ll}(a,b;\epsilon)=1$ if the geodesic crosses left lower face of the wedge and $\Theta_{cr}^{ll}(a,b;\epsilon)=0$ otherwise.

  Function $\Theta_{cr}^{rl}(a,b;\epsilon)=1$ if the geodesic crosses right lower face of the wedge and $\Theta_{cr}^{rl}(a,b;\epsilon)=0$ otherwise.

  Function $\Theta_{ncr}^{lu}(a,b;\epsilon)=1$ if the geodesic does not cross left upper face of the wedge and $\Theta_{ncr}^{lu}(a,b;\epsilon)=0$ otherwise.

 Function $\Theta_{ncr}^{ru}(a,b;\epsilon)=1$ if the geodesic does not cross right upper face of the wedge and $\Theta_{ncr}^{ru}(a,b;\epsilon)=0$ otherwise.

\begin{figure}[h]
    \centering
        \includegraphics[width=10cm]{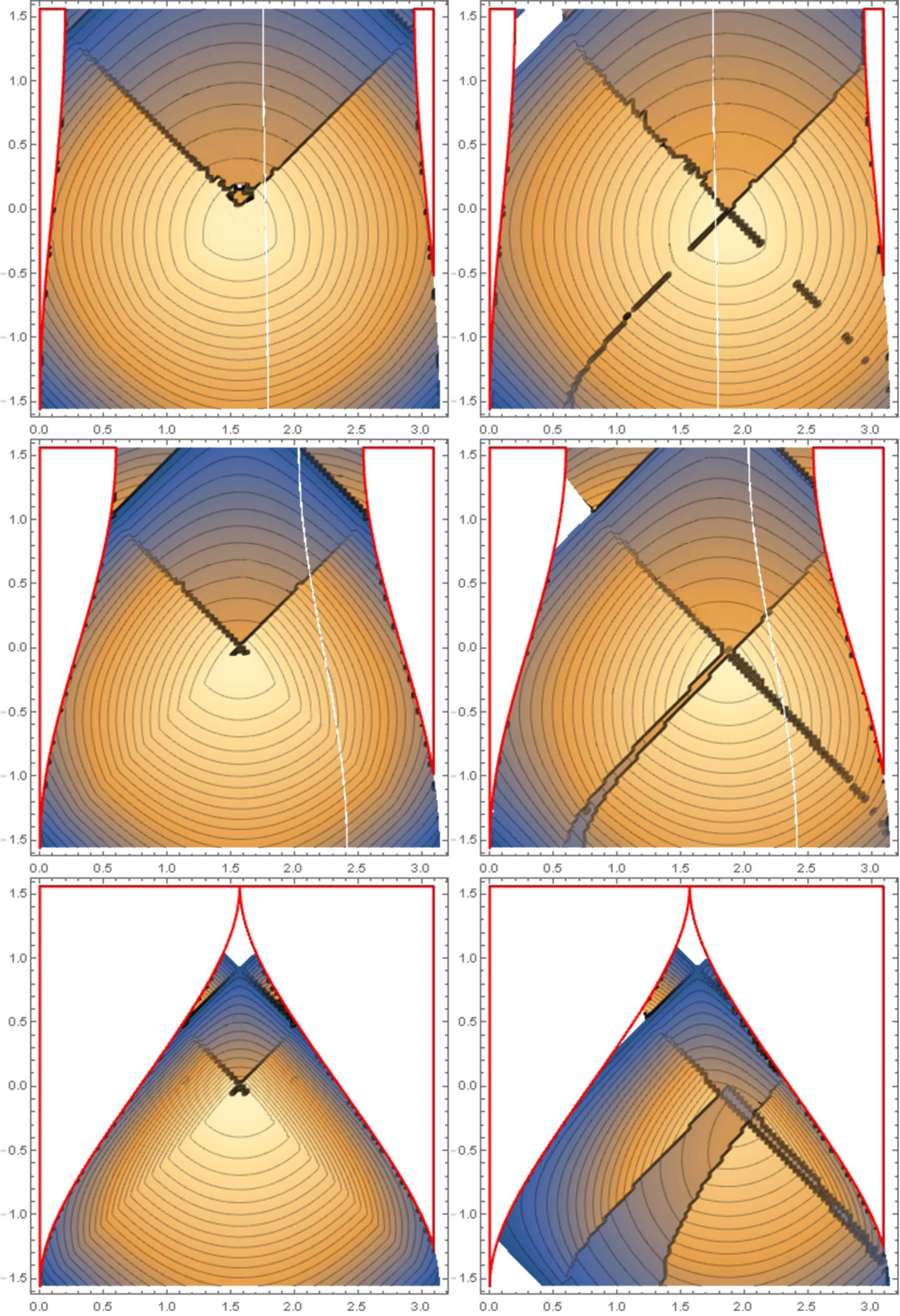}
       \caption{ Dependence of the inverse correlators in the case of two massless particles collision. In this case  $\phi_1=\frac{3\pi}{2}$ (left column), $\phi_1=5$ (right column) and $t_1=0$, conformal dimension $\Delta=1$ for each plot. Parameters for each plot are $\epsilon=0.1,0.3,0.78$ from top to down. Red curves correspond to boundaries of two dead zones, black curves are boundaries of discontinuities separating pulses. }
    \label{TPO1}
\end{figure}

\begin{figure}[h]
    \centering
        \includegraphics[width=10cm]{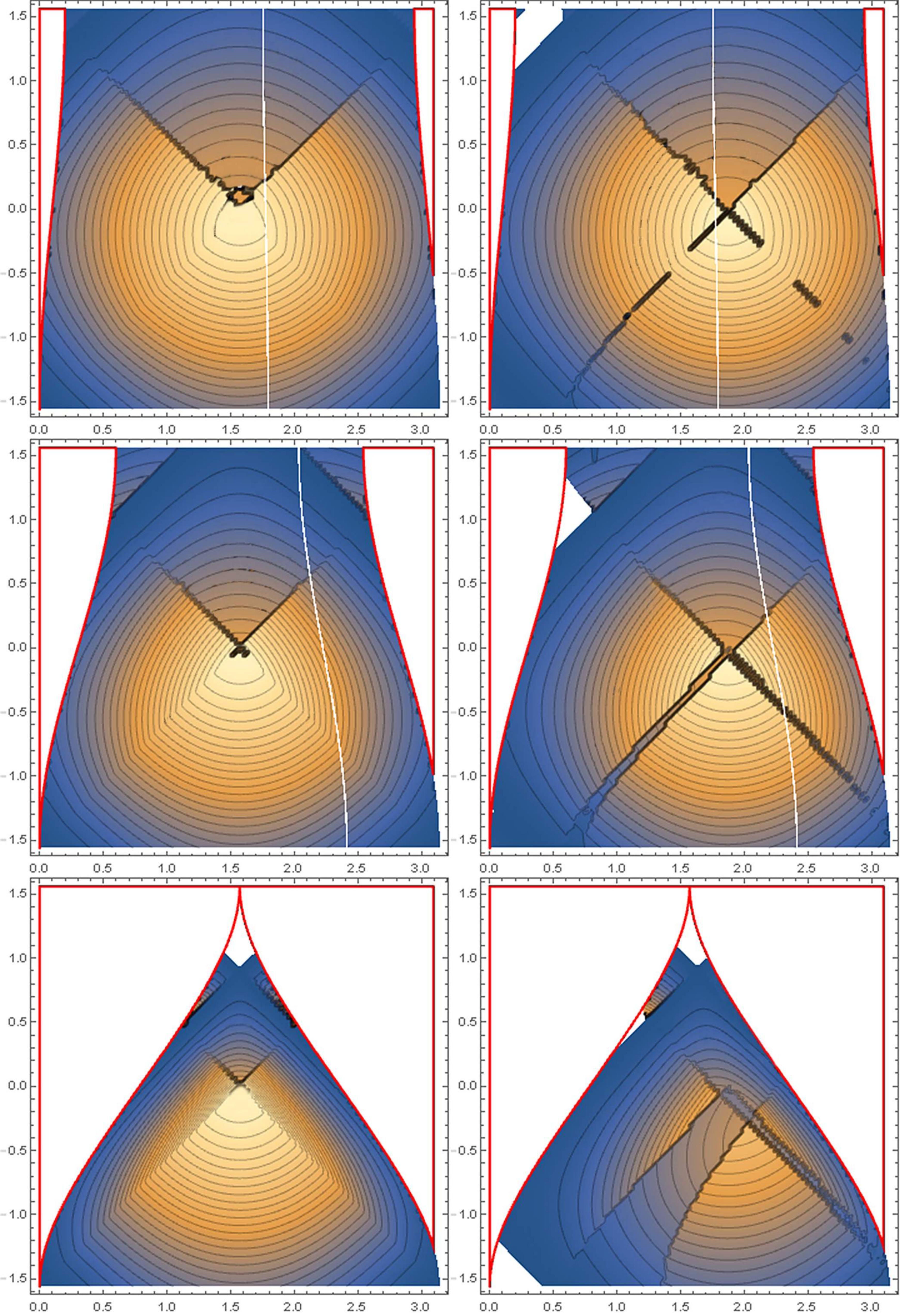}
        \caption{Dependence of the inverse correlators in the case of two massless particles collision. In this case  $\phi_1=\frac{3\pi}{2}$ (left column), $\phi_1=5$ (right column) and $t_1=0$, conformal dimension $\Delta=2$ for each plot. Parameters of collision for each plot are $\epsilon=0.1,0.3,0.78$ from top to down. Red curves correspond to boundaries of two dead zones, black curves are boundaries of discontinuities separating pulses }
    \label{TPO12}
\end{figure}

In Fig.\ref{TPO1} and  Fig.\ref{TPO12} we present the dependence of the inverse correlators on the boundary in the case of two massless particles in the bulk. Here $\phi_1=3\pi/2$ and $\phi=5$.
In all these plots we see two pulse zones coming from each boundary. We see asymmetry in the right columns of Fig.\ref{TPO1} and Fig.\ref{TPO12}. Asymmetry is related with the asymmetrical position of the point $\phi_1$ with respect to the collision line.  In some time these two pulses collide, forming another structure. Note, that the asymmetry is conserved during the collision process.

\begin{figure}[h]
    \centering
   \includegraphics[width=8cm]{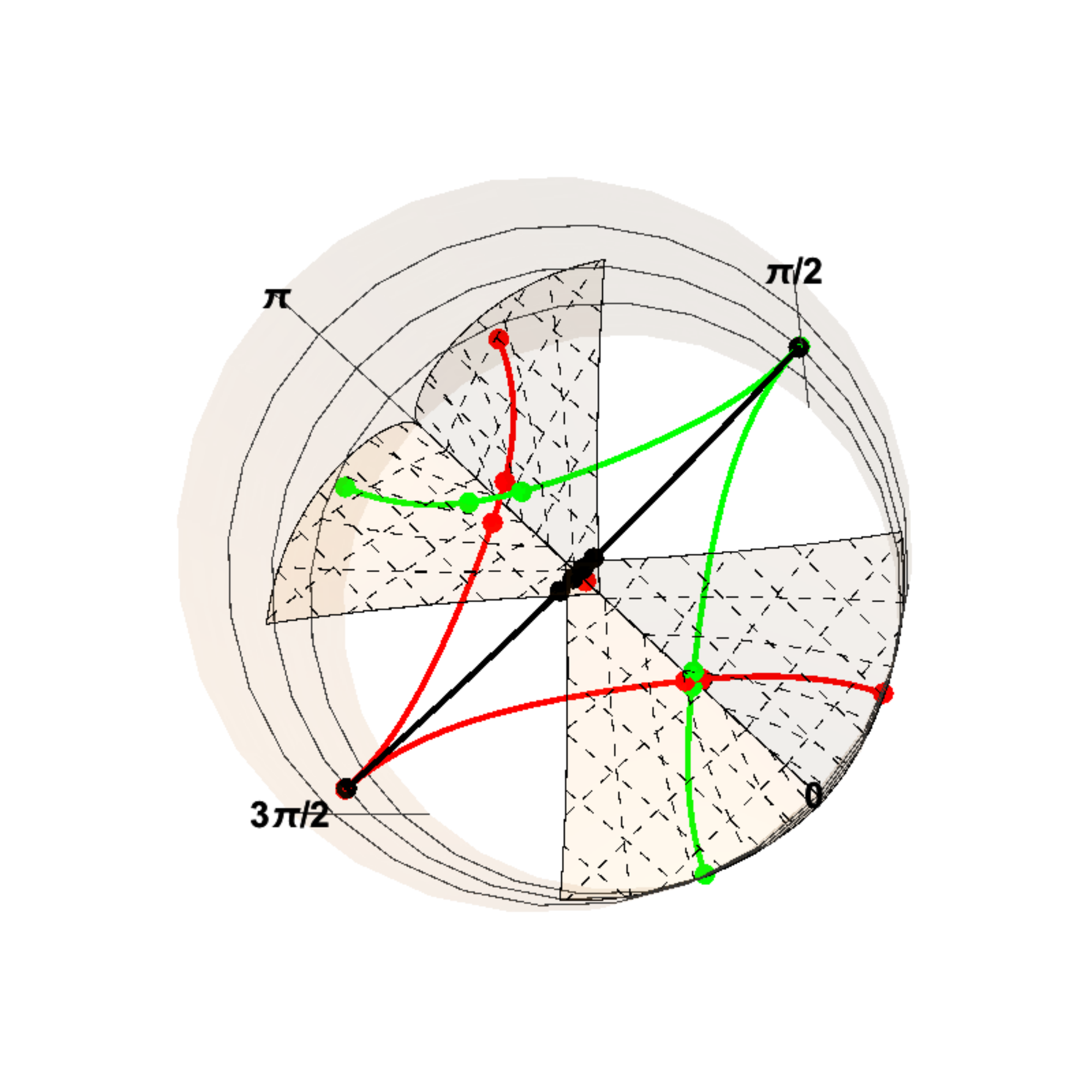}
            \caption{ Illustration of multiple geodesics configuration for two certain points (the black curve endpoints) on the boundary. Black curve is the basic geodesic, red geodesics pass through the left and right wedge faces, green geodesics are their image under the isometry action (here we take $\epsilon=\pi/4 $ for both defects).}\label{double-w}
           \label{double-w}
    \end{figure}

\newpage

\section{Holographic Entanglement Entropy Calculation}
\subsection{One massless defect.}
In this section we calculate the HEE for spacelike intervals for different equal time sections of our background.
We fix equal time points on the opposite sides of the boundary of the $AdS_3$ space with respect to the
particle worldline, $t_1=t_2=t_0$. To probe the HEE we vary $t_0$ from $t_0=-\pi/2$ to $\pi/2$. For a static spacetime the HEE  \cite{Nishioka:2009un} equals\footnote{we omit all prefactors in the calculation of the HEE like the gravitational constant etc.} to the minimal renormalized length $L_{ren}(\phi_1,t_0,\phi_2,t_0)$ of the geodesic connecting two points on the boundary, $(\phi_1,t_0)$ and $(\phi_2,t_0)$,

\be\label{HEE}
S(\phi_1,\phi_2,t_0)={\mbox{min}}\,\, {\cal L}_{ren}(\phi_1,t_0,\phi_2,t_0) .
\ee

\begin{figure}[h]
    \centering
        \includegraphics[width=6cm]{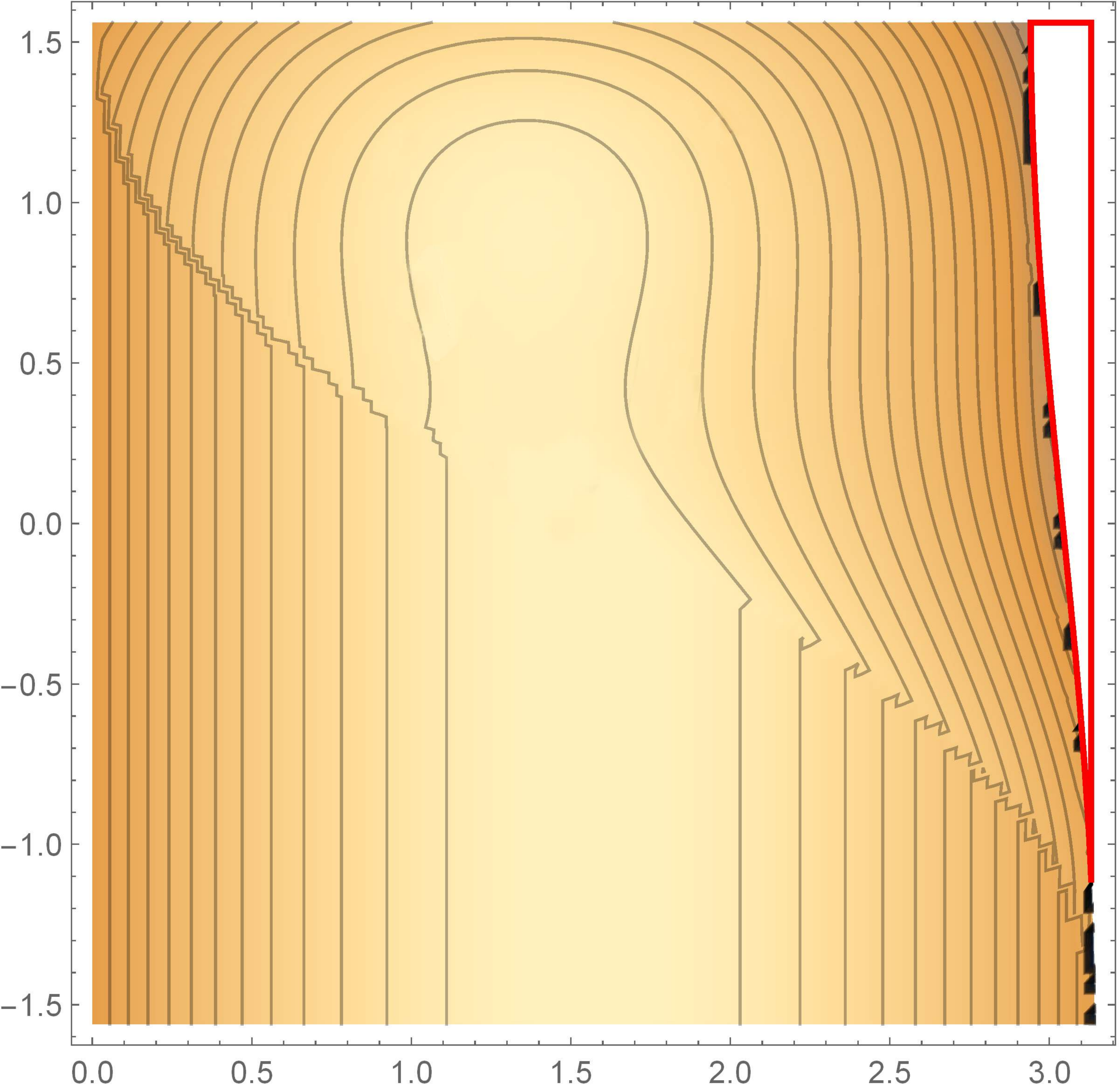}A.$\,\,\,\,\,$
        \includegraphics[width=6cm]{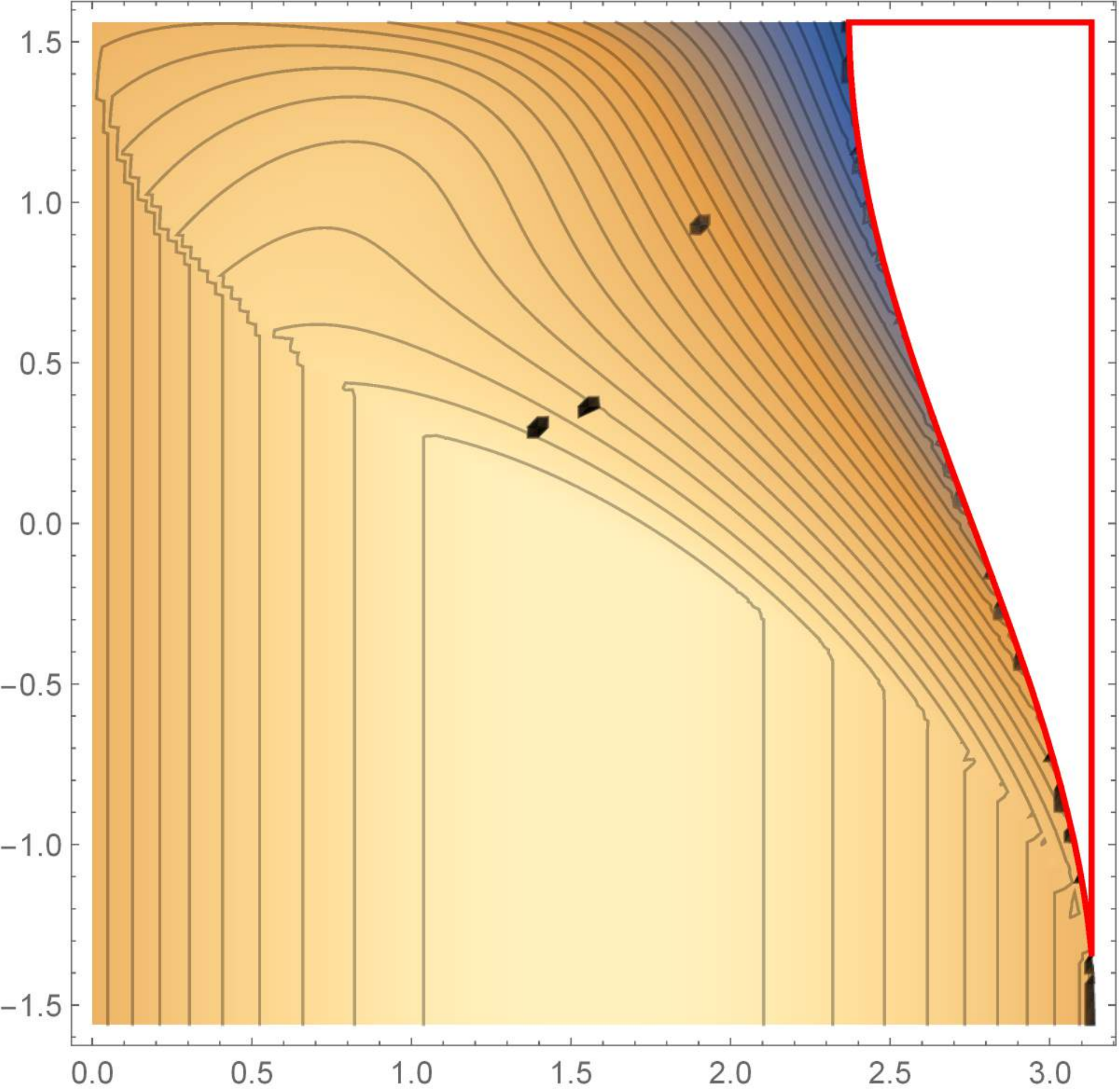}B.
     \caption{Dependence of the holographic entanglement entropy, calculated with the formula \eqref{HEE}, on $t_0$ and $\phi_2$. Here $\phi_1=\frac{3\pi}{2}$, $\epsilon=0.78$ (plot A) and  $\phi_1=\frac{3\pi}{2}, \epsilon=0.38$
            (plot B). $\phi_2$ corresponds to x-axis, $t_0$ to y-axis.}
                  $$\,$$
    \label{Fig:HEE1}
\end{figure}

As we can separate the time and space directions in the ADM formalism we can use formula \eqref{HEE} in this background.
In Fig.\ref{Fig:HEE1} we plot how the pulse structures described in the previous sections are probed by the HEE.
In Fig.\ref{Fig:HEE1} we plot the dependence of the  HEE  on $t_0$ and $\phi_2$ for $\phi_1$ being fixed  and different $\epsilon$. We see that similarly to  two-point functions,  there exists an expanding pulse-like zone propagating from the point on the boundary where the massless particle has been injected.

From Fig.\ref{Fig:HEE1}.A and Fig.\ref{Fig:HEE1}.B. we see that for some time the HEE remains constant, then it turns to a nonequilibrium regime. For larger intervals this transition to a nonequilibrium regime occurs faster.

\subsection{Two colliding particles}
In this subsection we probe the HEE evolution in the AdS$_3$ background deformed by two colliding massless particles.

Again, there are two different cases. The first one is when we  probe the HEE evolution for spacelike interval with endpoints  placed on the opposite sides of the boundary, with respect to the line of the collision, and the second one is when these points are on the same side with respect to the collision line. The configurations of the geodesics are the same as for calculations two-point correlators, but now we have to take into account only geodesics with
 minimal renormalized length.

\begin{figure}[h]
    \centering
        \includegraphics[width=6cm]{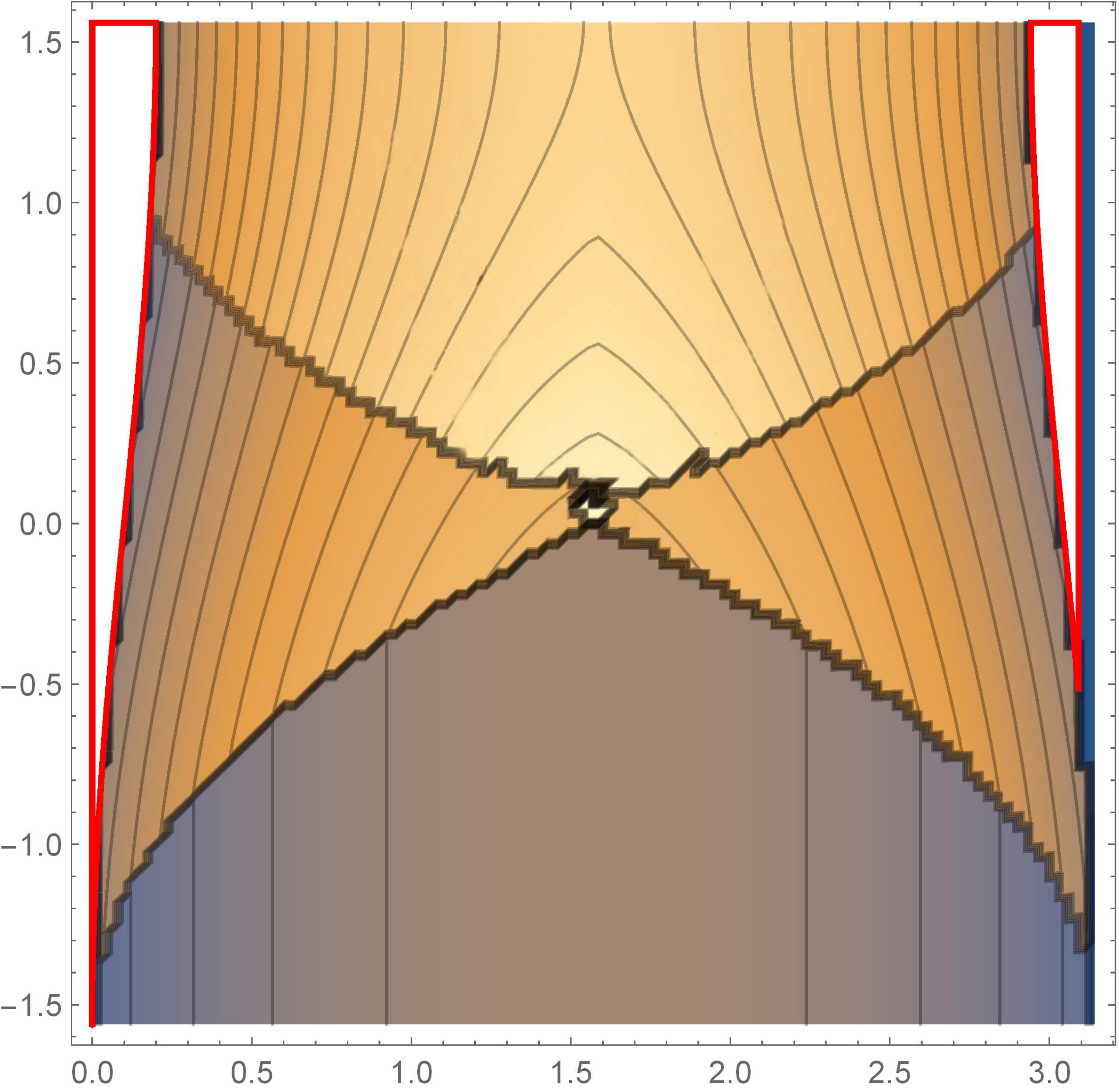}A.$\,\,\,\,\,$
        \includegraphics[width=6cm]{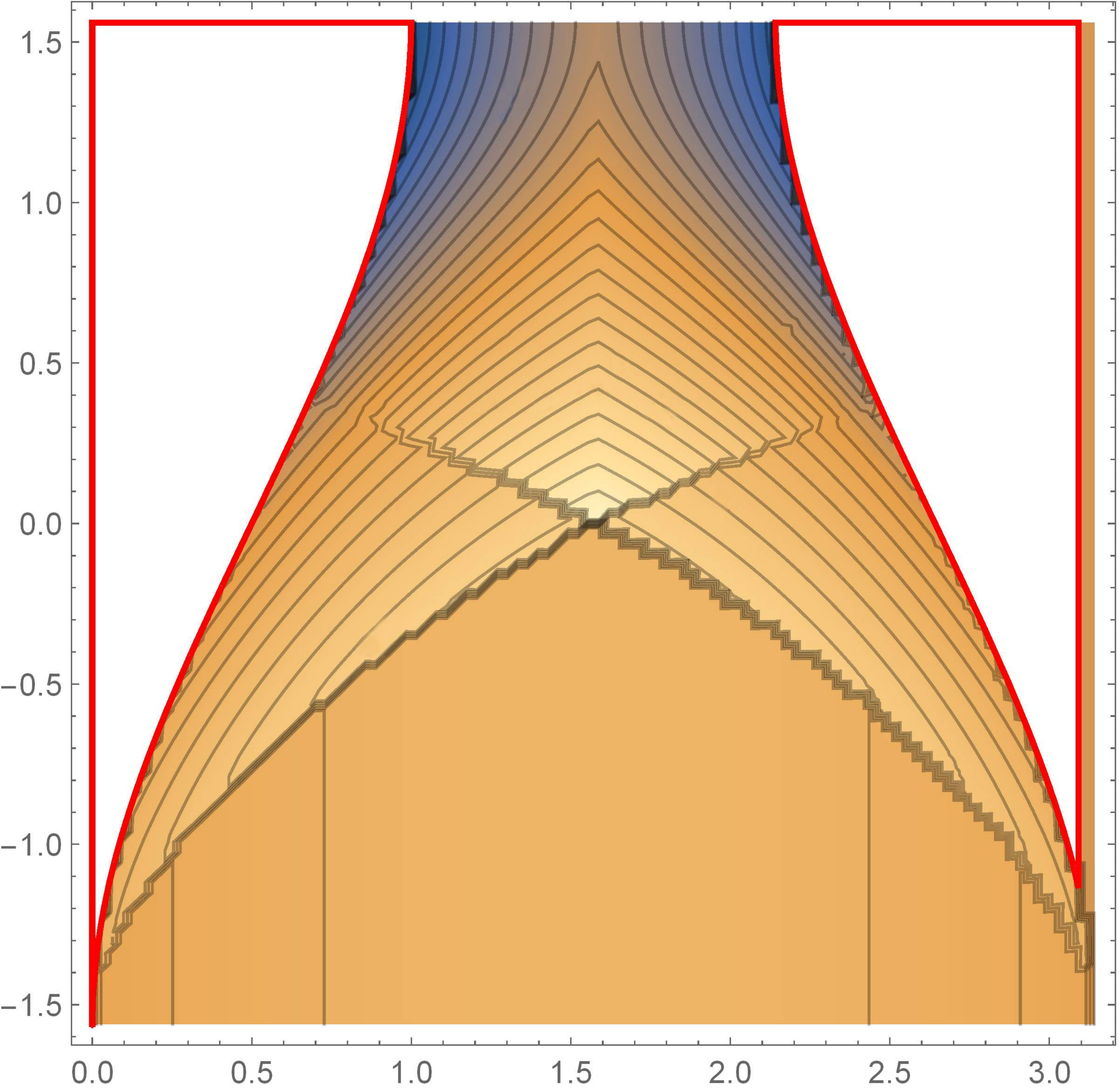}B.
            \caption{Dependence of the holographic entanglement entropy, calculated with the formula \eqref{HEE}, on $t_0$ and $\phi_2$  for different $\phi_1$ and $\phi_2$.  On the plot A. $\epsilon=0.1$ and $\phi_1=3\pi/2$. On plot B. $\epsilon=0.5$ and $\phi_1=3\pi/2$.
            $\phi_2$ corresponds to x-axis, $t_0$ to y-axis.}
    \label{Fig:HEE1-double}
\end{figure}

In Fig.\ref{Fig:HEE1-double}   we plot how the HEE probes the particles collision process,
we plot the dependence of HEE on $t_0$ and $\phi_2$ for different $\epsilon$ and fixed $\phi_1$. Here
 red curves correspond to the contracting boundary of the  living space. From  Fig.\ref{Fig:HEE1-double}A. we see wide zones, separated by discontinuities coming from each boundary. These zones rapidly grow, and from some moment the regime is changed again in all living space. In Fig.\ref{Fig:HEE1-double}.B the energy is large,
 $\epsilon=0.5$ and in Fig.\ref{Fig:HEE1-double}.B, $\epsilon=0.1$ and we see, that the size and rapidity of growth for these zones are relatively weak dependent   on the value of energy $\epsilon$.

$$\,$$
\newpage
$$\,$$
\newpage
$$\,$$
\newpage

\section{Conclusion}

In this paper we have considered the models of quantum field theory, dual to the $AdS_3$ space with one ultrarelativistic point-like particle and with two colliding ultrarelativistic point particles. In both cases the dual models
live on the spaces with varying sizes.
For these models, we have studied two-point correlation functions  within the geodesic approximation and the holographic entanglement entropy.  From numerical calculations we have seen that these models capture some  features of quantum systems under sudden quench and quantum systems with time-dependent volume.
We have shown that within geodesic approximation
the ultrarelativistic massless  defects
due to gravitational lensing of the geodesics  produce zone structure for correlators.  Non-stationary living spaces  produce   excitation waves, moving along  the boundaries  from the  quench points,  i.e. from the points on the boundary where the ultrarelativistic
particles is injected. The propagating pulse zone is localized near the ends of the wedge on the boundary.
There are also  intermediate zones separated by discontinuities which are  localized
    and  propagate along the living space with the constant speed.
HEE has also nontrivial zone structure. Two colliding massless defects produce more complicated zone structure for correlators and entanglement entropy.

 It worth to find out are these discontinuities in the correlator and entropy  artifacts  of
 the classical approximation that  we used, or they can be seen even in  a full solution through the scalar field equation. This question we suppose to investigate  in the separated study, that is started in \cite{AK}.

\section*{Acknowlegement}
We would like to thank Andrey Bagrov,  Mikhail Khramtsov, Maria Tikhanovskaya and Igor Volovich for useful discussions. This work was supported by the Russian Science Foundation (grant No. 14-11-00687).

  \end{document}